\documentclass[journal]{IEEEtran}
\usepackage{nopageno}
\usepackage{cite}
\usepackage{authblk}
\usepackage{graphicx}
\usepackage[utf8]{inputenc}
\usepackage{caption}
\usepackage{amsmath,bm}
\usepackage{tabularx}
\usepackage[linesnumbered,ruled,vlined]{algorithm2e}
\usepackage{algpseudocode}
\usepackage{xcolor}
\usepackage{amsfonts}
\usepackage[bottom]{footmisc}
\usepackage{subcaption}
\usepackage{comment}
\usepackage{multirow}
\usepackage{lscape}
\usepackage{pifont}
\usepackage{array}
\newcolumntype{Y}{>{\centering\arraybackslash}X}
\include{pythonlisting}
\usepackage{tikz}    % for encircle of text 

\ifCLASSINFOpdf

\else

\fi

\begin{document}
\title{An Optimized Privacy-Utility Trade-off Framework for Differentially Private Data Sharing in Blockchain-based Internet of Things}   
\author[1]{Muhammad Islam}
\author[2]{Mubashir Husain Rehmani}
\author[3]{Jinjun Chen}
%\affil[1]{Swinburne University of Technology, Hawthorn, VIC 3122, Australia} 
\affil[1,3]{Swinburne University of Technology, Hawthorn, VIC 3122, Australia}
\affil[2]{Munster Technological University, Rossa Avenue, Bishopstown, Cork, Ireland}
\renewcommand\Affilfont{\itshape\small}

\maketitle
\thispagestyle{empty}
\begin{abstract}
Differential private (DP) query and response mechanisms have been widely adopted in various applications based on Internet of Things (IoT) to leverage variety of benefits through data analysis. The protection of sensitive information is achieved through the addition of noise into the query response which hides the individual records in a dataset. However, the noise addition negatively impacts the accuracy which gives rise to privacy-utility trade-off. Moreover, the DP budget or cost $\epsilon$ is often fixed and it accumulates due to the sequential composition which limits the number of queries. Therefore, in this paper, we propose a framework known as optimized privacy-utility trade-off framework for data sharing in IoT (OPU-TF-IoT). Firstly, OPU-TF-IoT uses an adaptive approach to utilize the DP budget $\epsilon$ by considering a new metric of population or dataset size along with the query. Secondly, our proposed heuristic search algorithm reduces the DP budget accordingly whereas satisfying both data owner and data user. Thirdly, to make the utilization of DP budget transparent to the data owners, a blockchain-based verification mechanism is also proposed. Finally, the proposed framework is evaluated using real-world datasets and compared with the traditional DP model and other related state-of-the-art works. The results confirm that our proposed framework not only utilize the DP budget $\epsilon$ efficiently, but it also optimizes the number of queries. Furthermore, the data owners can effectively make sure that their data is shared accordingly through our blockchain-based verification mechanism which encourages them to share their data into the IoT system.         
\end{abstract}
\begin{IEEEkeywords}
IoT, Blockchain, differential privacy, data sharing, privacy-utility trade-off. 
\end{IEEEkeywords}
\IEEEpeerreviewmaketitle
\section{Introduction}
\label{sec:intro}
Internet of Things (IoT) has provided an opportunity for enhanced, intelligent, and smart services and applications in various domains such as smart health, smart cities, smart industry, intelligent transportation, and recommender systems \cite{IoTsurvey}. The backbone of these applications is the data collection on large scale from IoT devices and then mining it for beneficial trends and patterns which are further used in intelligent decision making. For instance, medical data collected in hospitals can be utilized to provide useful insights for practitioners and researchers \cite{IoTforsmarthealth}. Similarly, in a smart factory, the data collected from various machines and devices can be used for predictive maintenance \cite{IoTforsmartfact}. In intelligent transportation, the data collected from vehicles can be utilized for traffic control purposes \cite{IoTforsmarttransport}. Moreover, the data collected from vehicles, smart factories and hospitals can be used for smart urban living through smart cities \cite{IoTinsmartcity}. However, it is very common that the data owner’s sensitive and identification information may leak during the analysis of the data. As a result, data owners are often reluctant to share their data \cite{IoTuserpriv1, IoTuserpriv2, IoTuserpriv3}.\\  
\begin{figure*}[htp]
\centering
\begin{subfigure}[t]{0.45\linewidth}
\centering
\includegraphics[width=\linewidth]{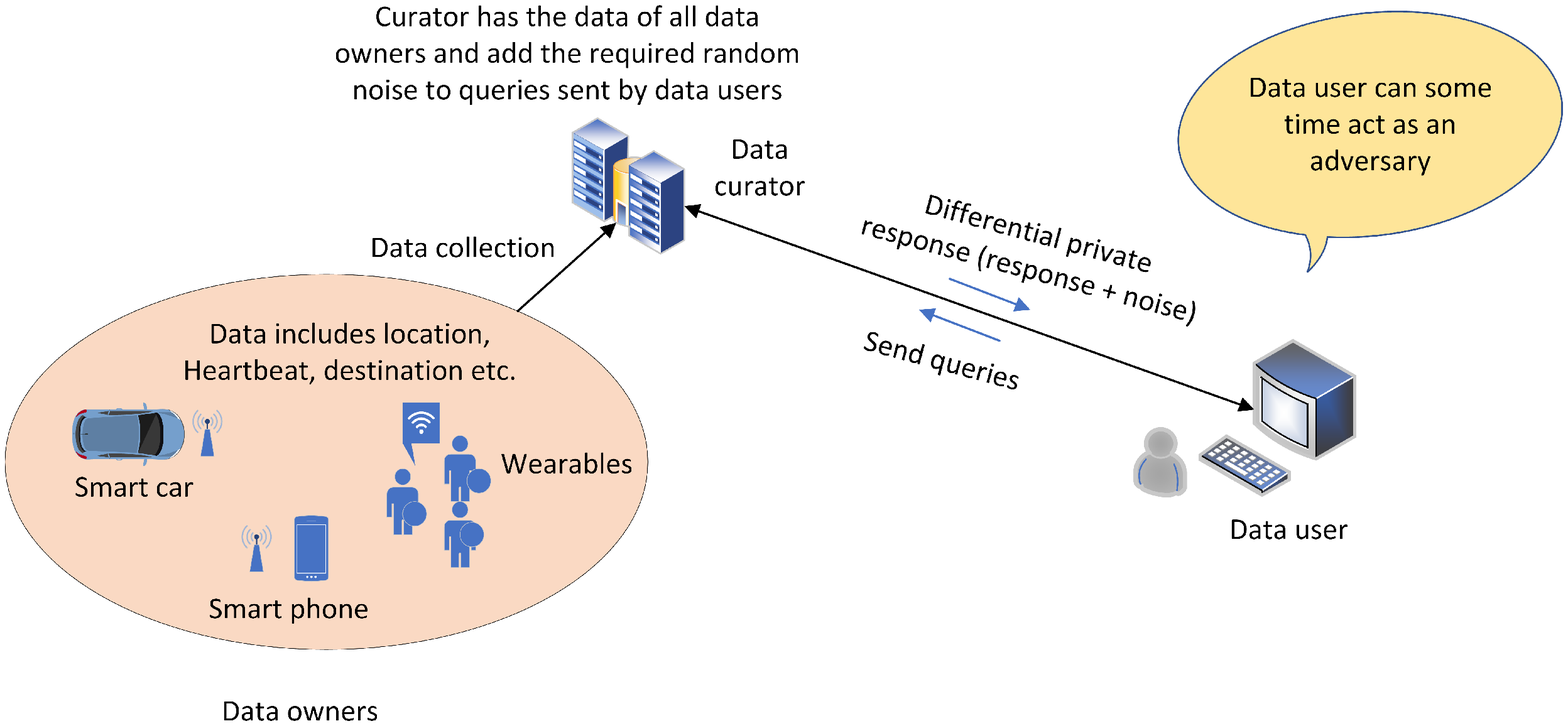}
\caption{ }
\label{subfig:sample_model}
\end{subfigure} \hspace{10mm}
\begin{subfigure}[t]{0.45\linewidth}
\centering
\includegraphics[width=\linewidth]{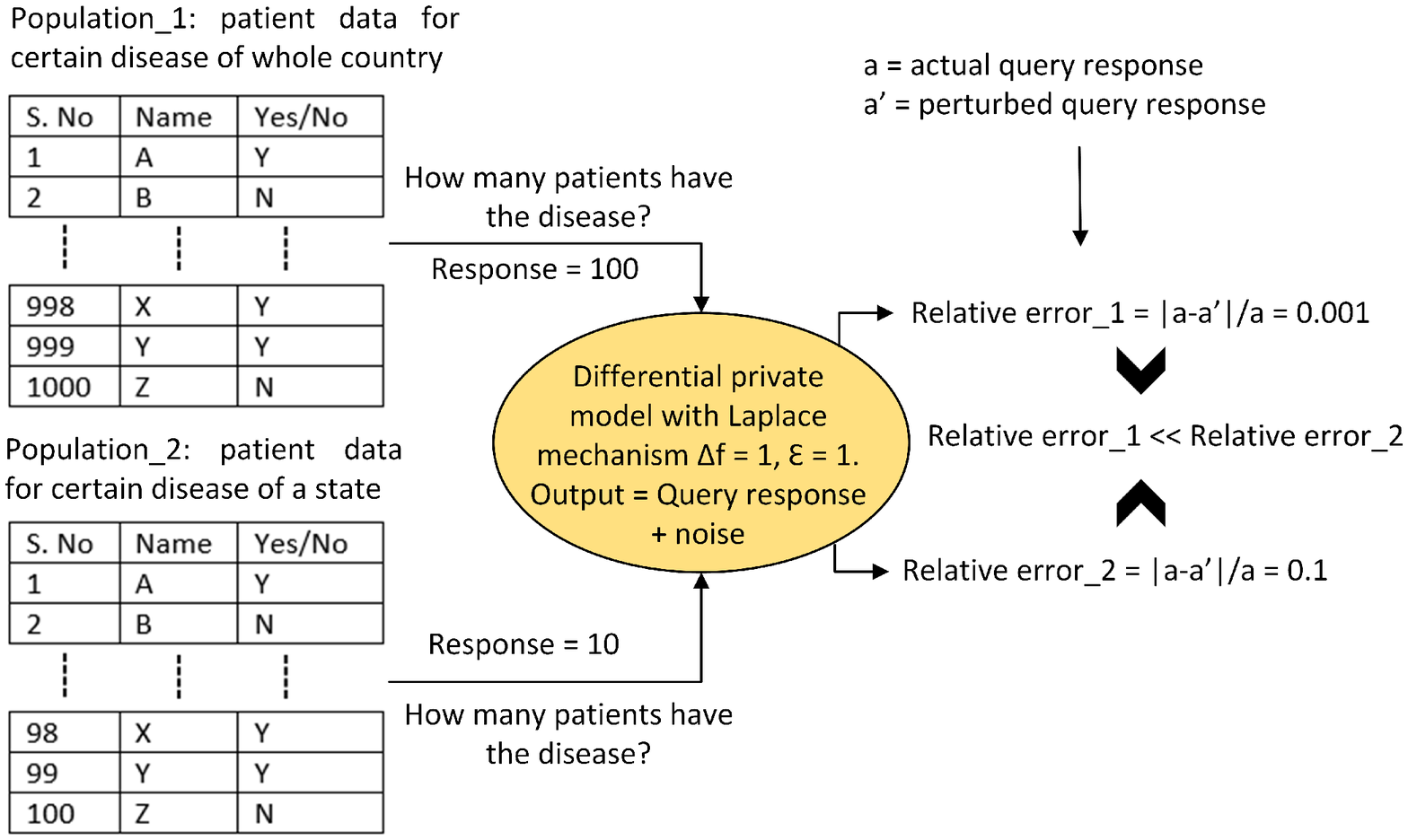}
\caption{ }
\label{subfig:rel_err_vs_pop}
\end{subfigure}
\caption{(a) Differential private data sharing in IoT, (b) illustration of relative error vs population size.}
\label{fig:fig_1}
\end{figure*} 
Recently, differential privacy got popularity in the context of private data sharing in IoT \cite{differentialpubsurvey}. The main principle of differential privacy is that it hides an individual record in a group of records or dataset while calculating the aggregated results \cite{Dwork2008}. The aggregated results could be either published at once or shared in the form of responses to queries sent by data users. In this work, we consider the data sharing through queries because it is more practical than the other one \cite{RLbasqueropt}. In this case, the data sharing model consists of a set of data owners, a data curator, and a data user (or set of data users). A scenario of the data sharing model in IoT is shown in Fig \ref{fig:fig_1}(a). In Fig. \ref{fig:fig_1}(a), the curator is trusted but the data user can act as an adversary due to which the sensitive information can be leaked regarding the data owners. To protect this leakage, the curator inserts a random noise before sharing the response with the data user. More random noise added into the query response means high privacy preservation and vice versa. However, the random noise negatively impacts the accuracy of the query response. Furthermore, a trade-off between privacy and utility or accuracy is developed, i.e., increased privacy will result in less accuracy and vice versa \cite{gambasedpriv_utility}. It is to be noted here that our previous work \cite{islamtransparency} presented transparency-privacy trade-off problem which is different from privacy-utility trade-off problem. Moreover, the current work focuses on optimizing the trade-off and increasing the number of queries under a given differential privacy budget. \\
Similarly, another drawback of differential privacy is that the privacy budget or privacy cost denoted as $\epsilon$ accumulates for sequential queries due to the sequential composition of differential privacy \cite{differentialpubsurvey}. Consequently, a given privacy budget allows small number of queries under the constraint of differential privacy. Due to these drawbacks, differential private models cannot be adopted on large scale in IoT-based applications. In this context, various studies suggested innovative techniques to solve the above-mentioned problems. For instance, game theoretic models were adopted to solve the privacy-utility trade-off problem by selecting suitable values of privacy budget in \cite{gambasedpriv_utility, dist-class-priv-acc, privacy-dist}. Furthermore, to satisfy both data owners and data users, reinforcement-based and heuristic-based approaches were adopted to optimize the number of queries \cite{RLbasqueropt, adoptive_person_DP, util_awar_gen_framework}. Moreover, in \cite{trackbudget}, a mechanism was introduced to efficiently utilize the privacy budget and blockchain was adopted to satisfy the data owners regarding the utilization of their data. \\
However, the above-mentioned works have used a fixed privacy budget allocation approach and failed to consider the relationship of population size (dataset size) and accuracy of the query response for various query functions such as count, average, median and mode. For instance, Fig. \ref{fig:fig_1}(b) explains how the population size impacts the accuracy of the count query response \cite{dp_pop_size}. For example, the population\_1 represents the medical records for a whole country and the population\_2 represents the medical records for people of a state or specific area code. Also, in real-world scenarios, data users are not always interested in query evaluation over the whole population (dataset), and they need query evaluation only on a specific portion of the population (dataset). It is evident from Fig. \ref{fig:fig_1}(b) that for the same privacy budget and query type, the Relative error\_1 for population\_1 with size 1000 is much smaller than the Relative error\_2  for population\_2 with size 100. Furthermore, relative error and accuracy are inversely proportional, therefore, a high value of relative error means low accuracy and vice versa. In other words, to get same level of accuracy of the query response over the two populations, the privacy budget for population\_1 needs to be less than the privacy budget for population\_2. However, the existing mechanisms treat each data user in a uniform manner and don't consider the population size. Consequently, the query is evaluated with the same allocated budget for each population which satisfy both data owner and data user, but the privacy budget is wasted due to which its allocated value is exhausted quickly. As a result, the total number of queries that can be answered reduces. Apart from this, to avoid the centralized authority from processing and collecting the data, local differential privacy model is adopted, however, it significantly reduces the accuracy of the aggregated results. As a result, to get accurate aggregated results with transparency in the operations of the centralized curator, more efficient and enhanced privacy preserving models are needed. \\
Therefore, this paper proposes an optimized privacy-utility trade-off framework for differentially private data sharing in IoT (OPU-TF-IoT) to address all the above-mentioned problems. The novelty of the proposed framework is that it uses an adaptive approach to utilize the privacy budget by considering a new metric of population size and priority along with the data user's query to optimize the number of queries. Similarly, an algorithm is proposed to avoid the waste of privacy budget due to the uniform treatment of data users. Moreover, it reduces the privacy cost accordingly while satisfying the needs of the data owners and data users. Finally, inspired from the work in \cite{trackbudget}, a verification mechanism using blockchain technology is proposed for data users to verify that the data is shared accordingly. The main contributions of our work are as following.  
\begin{itemize}
\item An optimized privacy-utility trade-off framework for IoT-based applications (OPU-TF-IoT) is proposed which uses an adaptive approach by considering the new metric of population size along with the query to optimize the number of queries while satisfying both data owners and data users.
\item An algorithm is proposed to reduce the privacy cost by avoiding its waste due to uniform treatment of data users through heuristic search thus utilizing the saved privacy budget for more query responses.
\item A new blockchain-based mechanism is proposed through which data owners can verify the utilization of privacy budget or cost which increases their satisfaction on the data sharing system.  
\item A comprehensive comparison is presented using real-world datasets to verify the improvement of the proposed framework (OPU-TF-IoT) over the traditional differential privacy model and other state-of-the-art mechanisms in terms of optimized privacy-utility trade-off. 
\end{itemize}
The rest of the paper is organized according to the following sequence. Section II presents the relevant works from the literature. Similarly, Section III presents the proposed framework in detail. Section IV presents the performance evaluation and comparison. Finally, Section V concludes the work.   
\section{Literature review}
\label{sec:let_rvw}
Previously, various techniques were adopted to efficiently utilize the differential privacy budget in order to optimize the privacy-utility trade-off. For instance, in \cite{trackbudget}, a blockchain-based mechanism has been proposed to record each query, its type, and the privacy budget utilized in generating the perturbed response. Afterwards, it checks the new incoming queries against the record, and if it successfully finds it then the previous response is returned instead of utilizing new privacy budget. In this way, the privacy budget is utilized in an efficient manner.  However, the data users have been treated in uniform manner and their priorities have been ignored which results in waste of privacy budget. Furthermore, if the number of repeated queries is small then this technique is not successful, and it acts like the traditional differential privacy model. \\ 
Similarly, another the work in \cite{RLbasqueropt} have introduced the mechanism of batch queries which acts similar to the previous technique of \cite{trackbudget}. Furthermore, a heuristic search algorithm has been used to find a suitable setup where the data curator can satisfy maximum number of data users while maintaining the privacy preservation level of data users. Moreover, to reduce the complexity of the heuristic algorithm reinforcement learning has been adopted which significantly improves the search time. To this end, their proposed approach performs better than the traditional differential privacy model. However, disjoint dataset has been considered despite that the records in real-world scenarios are corelated. Furthermore, single query has been considered while multiple queries have been ignored which limits its applicability in IoT applications. As opposed to the previous two techniques, an adaptive approach has been used to optimize the privacy-utility trade-off problem in \cite{adoptive_person_DP}. The proposed approach adopts suitable noise generating algorithms based on distribution of data, query functions and privacy settings which improves the privacy-utility trade-off. However, theoretically it is not practical to find an optimized threshold for the sampling of tuples which maintain the same utility of data. Furthermore, the priorities of data users have been ignored.\\
Another similar work in \cite{util_awar_gen_framework} have proposed a general framework for location privacy preservation. The main idea is to utilize different noise scale to each point in a trajectory of movement which guarantees the utility. However, it also adopts the traditional differential privacy model for utilization of privacy budget and efficient utilization of privacy budget is not the primary focus. Furthermore, the model is specifically developed for location privacy preservation and thus cannot be adopted for other scenarios. Similarly, in \cite{gambasedpriv_utility}, the problem of correlated records has been discussed. More specifically, the proposed model considers that the user privacy is not only affected by its own choice of privacy budget, but it is also affected by the choice of privacy budget of its neighbors. To this end, it improves the privacy preservation of individuals in correlated databases. However, the optimization of privacy-utility trade-off has been ignored.\\
Apart from this, in \cite{privacy-dist}, a total variation distance has been adopted to measure the privacy leakage. The proposed approach showed that the optimal privacy-utility trade-off problem can be solved by a standard linear program. However, the proposed model is very general, and it does not consider the relationship between population size and accuracy of the data. Consequently, although, it solves the privacy-utility trade-off problem, however, the mechanism to avoid waste of privacy budget is missing. In \cite{contract-theoratic}, privacy has been modelled as goods to be sold i.e., between data owners and data collectors. A contract theoretic approach is then proposed in which data collector deals with the privacy-utility trade-off. A contract between the parties is signed which describes how much prices should the data owner receive for a certain level of privacy preservation. In this way, the data collector takes better decision whether higher a higher utility is needed or less price is to be paid for providing higher guarantee of protecting privacy of data owners.
\begin{table}[!b]
\caption{Key notations and its description}
\begin{center}
\resizebox{0.9\linewidth}{!}{%
\begin{tabular}{ll}%{tabular}{|c|c|c|c|c|c|}
\hline
Notation & Meaning \\
$\epsilon$ & Differential privacy budget \\
$\epsilon_{sut}$ & Suitable privacy budget \\
$\epsilon_{t}$ & Total privacy budget \\
$\epsilon_{def}$ & Default privacy budget $0 < \epsilon_{def} < \epsilon_{t}$ \\
$\mu$ & Mean for Laplace distribution \\
$\lambda$ & Laplace scale \\
${\bigtriangleup}f$ & Sensitivity \\
HSA & Heuristic search algorithm \\
\textbf{C} & Data curator \\
\textbf{O} & Set of data owners \\
\textbf{U} & Set of data users \\
$\bm{\epsilon}$ & Set of the desired $\epsilon$ values from data users \\
$q_{i}$ & $i^{th}$ query \\
$q^{'}_{i}$ & $i^{th}$ query response \\
$T_{n, m}$ & Data table \\
$A^{i}_{req}$ & Required accuracy by $i^{i}$ data user \\
$A^{i}_{act}$ & Calculated/actual accuracy of $i^{th}$ query \\
$F$ & Query function \\
$N$ & Query type \\
$r^{i}_{err}$ & Relative error in the $q^{'}_{i}$ \\
$\Upsilon_{i}$ & Numerical value of the query $q_{i}$ \\
$\tau$ & Tolerance coefficient \\
$\eta$ & Decrement factor \\
$\rho$ & Minimum no of satisfied data users \\
\hline
%\end{tabular}
\end{tabular}}
\label{tab:notation}
\end{center}
\end{table}

\begin{figure*}[!htp]
\begin{center}
\includegraphics[width=12cm]{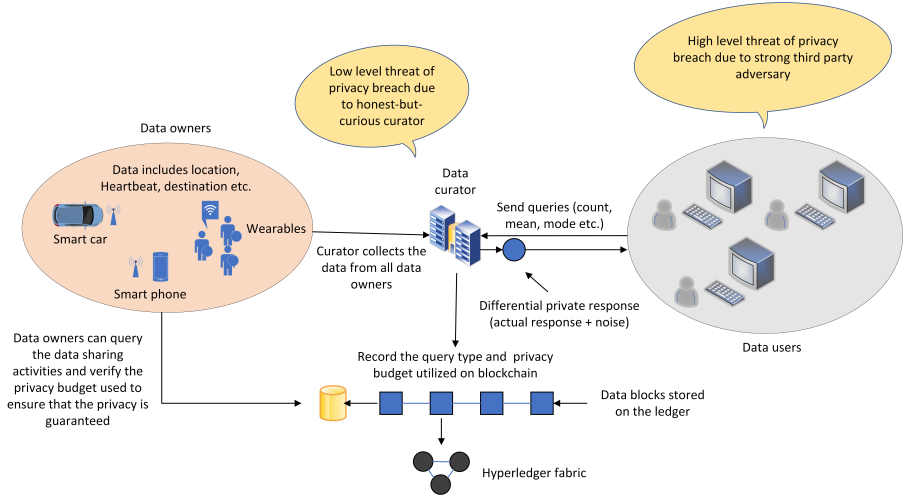}
\caption{Demonstration of query-response mechanism with the recording of query on blockchain.}
\label{fig:system_model}
\end{center}
\end{figure*}
In \cite{dist-class-priv-acc}, the problem of privacy-accuracy trade-off has been discussed in the context of distributed data mining. The selection of privacy level by individual users impacts the accuracy of data for data classifier or mediator. Similarly, a different game model is adopted to represent the interaction among users in which a user cannot observe the privacy budgets of others. The existence of satisfaction equilibrium (SE) is then proved in which each user is satisfied in their individual constraints. However, the focus of these works is not related to utilization and maximizing the number of queries. Similarly, the work in \cite{ppass2021} has proposed a generic model for selecting a suitable privacy preservation mechanism based on the dimensions or type of dataset. Furthermore, fuzzy logic has been used to get a fuzzy index (FI) which decides which privacy mechanism to be selected. However, an explicit privacy-utility trade-off, optimization and increase the number of queries, and avoid the waste of privacy budget are missing. Furthermore, the calculation of FI is costly in terms of computation, hence, not scalable. \\
Therefore, due to the above-discussed limitations of the existing mechanisms, the differential private data sharing in IoT still needs further improvement. More specifically, the problems of optimization of privacy-utility trade-off, waste of privacy budget due to the uniform treatment of data users, and lack of a verification mechanism for data owners to track the privacy budget and data sharing activities still open to the research community. To this end, in this paper, we present a solution to the above-mentioned problems though our optimized privacy-utility framework (OPU-TF-IoT).   
\section{Proposed Work: OPU-TF-IoT}
\label{sec:prop_work}
In this part of the paper, we present our proposed framework in detail. Firstly, to set-up the background, the preliminaries section presents the basics of differential privacy model and blockchain. Afterwards, system model, adversary model, and the proposed heuristic search algorithm (HSA) are presented. Moreover, throughout this paper, the word dataset is used to represent a population and accuracy is used to represent the utility of the data. Similarly, $\epsilon$ denotes the privacy budget or cost. Other notations used in the paper are summarized in the Table \ref{tab:notation}.  
\subsection{Preliminaries}
\label{subsec:prel}
\subsubsection{Differential privacy}
\label{subsubsec:dp}
C. Dwork for the first time introduced differential privacy for statistical databases \cite{Dwork2008}. It is based on the principle which states that the output of an algorithm applied to a dataset will not change in a significant way by adding or removing a single record from the dataset. The formal definition of differential privacy mentioned in \cite{Dwork2008} is given as following:\\\\
\textbf{Definition 1.} \textit{A randomized function Z satisfies $\epsilon$-differential privacy if for all datasets $D_{i}$, $D_{j}$ which differs in one record, and for all $S \subseteq Range(Z)$, the following holds \cite{Dwork2008}:}
\begin{align}
P[Q(D_{i}) \in S] \leq e^{\epsilon} \times P[(Q(D_{j})\in S] && \text{(by \cite{Dwork2008})} \label{eq:eqdp}
\end{align}
Where Range(Z) is the range of all possible outputs of function Z, $\epsilon$ is differential privacy budget such that $\epsilon > 0$, and $D_{i}$, $D_{j}$ are the two neighboring databases such that $D_{j}$ is generated by removing or adding a single record of the data from $D_{i}$ and vice versa.
Furthermore, the maximum difference between query answers over $D_{i}$ and $D_{j}$ is known as \textit{sensitivity} which is denoted as ${\bigtriangleup}f$. The \textit{sensitivity} depends on the type of query function. For instance, in case of count queries, the maximum difference between query responses calculated over $D_{i}$ and $D_{j}$ is 1. Therefore, mathematically it can be written as following:  
\begin{align}
{\bigtriangleup}f = {| f(D_{i})-f(D_{j}) |}_{1} && \text{(by \cite{Dwork2008, dwork2014algorithmic})} \label{eq:sensitivity}
\end{align}
Furthermore, in literature, two popular mechanisms have been used to implement differential privacy which are (i) Laplace mechanism, and (ii) Exponential mechanism \cite{differentialpubsurvey}. We use Laplace mechanism because it is suitable for numerical queries. The Laplace distribution function is given as following:
\begin{align}
Lap(x, \mu, \lambda) = \frac{1}{2\lambda}e^{\frac{-|x-\mu|}{\lambda}} && \text{(by \cite{differentialpubsurvey})}\label{eq:lap_dist}
\end{align}
Where $\lambda$ and $\mu$ are the Laplace scale and mean for the Laplace distribution, respectively. Furthermore, $\lambda = \frac{{\bigtriangleup}f}{\epsilon}$, and $x \in \mathbb{R}$. \\
Apart from this, two composition theorems have been discussed in the context of differential privacy which are given below \cite{differentialpubsurvey}.
\paragraph*{Theorem 1 (Parallel composition)} for a set of privacy preserving mechanisms $\textbf{M} = \{M_{1}, M_{2}, M_{3}…M_{m}\}$, if every mechanism $M_{i}$ satisfies differential privacy equivalent to $\epsilon_{i}$ on the disjoint subsets of the dataset $D_{i}$ then \textbf{M} will satisfy differential privacy equivalent to max-$\epsilon_{i}$ \cite{differentialpubsurvey}.
\begin{figure*}[!thp]
\begin{center}
\includegraphics[width=13cm]{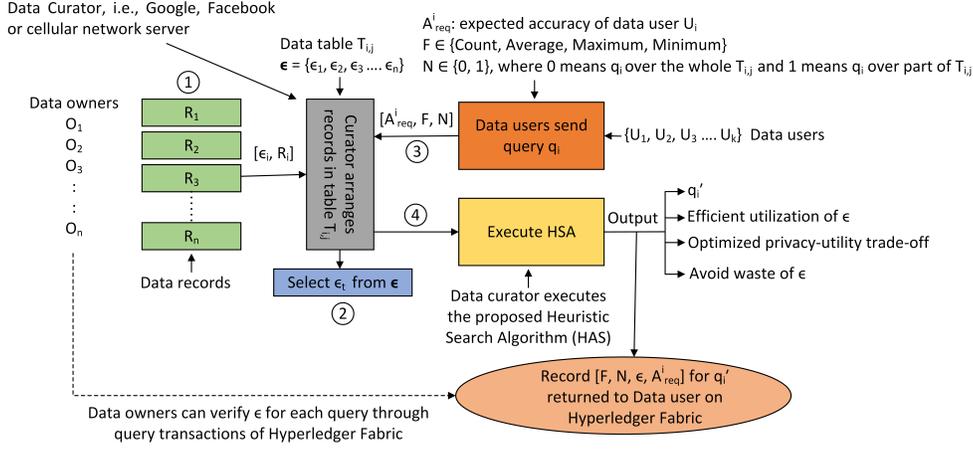}
\caption{Overview of the proposed framework (OPU-TF-IoT).}
\label{fig:fram_design}
\end{center}
\end{figure*}
\paragraph*{Theorem 2 (Sequential composition)} for a set of privacy preserving mechanisms $\textbf{M} = \{M_{1}, M_{2}, M_{3}…M_{m}\}$, if every mechanism $M_{i}$ satisfies differential privacy equivalent to $\epsilon_{i}$ on the same dataset $D_{i}$ then \textbf{M} will satisfy differential privacy equivalent to $\sum_{i=1}^{m} \epsilon_{i}$ \cite{differentialpubsurvey}. 
\subsubsection{Blockchain}
\label{subsubsec:hf}
Satoshi Nakamoto was the first who gave the concept of blockchain in 2008 for virtual currency \cite{nakamoto2008}. Blockchain is a distributed ledger which keeps the record of each transaction in the form of cryptographically connected data blocks. In a distributed environment, it solves the problem of lack of trust among the participants through transparency, traceability, and verification of each transaction \cite{anomalyinBC, compriot2020}. Currently, it has been adopted in various network scenarios to increase the transparency of operations, avoid frauds, and enable tracking and provenance. Every change, transaction, and network activity are verified through its distributed consensus mechanism which increases the trust among the participants of the network. \\  
As a result, we adopt blockchain to address the lack of transparency in the context of data processing, collection, and sharing by the centralized curator. In this way, we leverage the high accuracy of aggregated results calculated over the data in the context of a centralized data curator whereas the processing and sharing of data is made transparent to the data owners. In this work, we adopt Hyperledger fabric.  Furthermore, the transaction processing is fast as compared to other types of blockchain which results in high throughput \cite{hyper}. 

\subsection{System Model of OPU-TF-IoT}
\label{subsec:system_model}
The proposed system model is presented in Fig. \ref{fig:system_model}. Furthermore, the detailed description of each component is given as following. 
\subsubsection{Centralized Data Curator} 
\label{subsub:curator}
A data curator denoted as \textbf{C} collects data from the data owners. For example, it could be a server of Facebook, Google, or Cellular network. Furthermore, it has its own database in which the collected data is recorded. Furthermore, the data is shared with third parties or governmental agencies for analysis. In this context, data curator and data owners agree on a maximum value of privacy budget known as total privacy budget $\epsilon_{t}$ which is then used to perturb the data before sharing to ensure the privacy protection of sensitive information of individual data owners. 
\subsubsection{Data Owner} 
\label{subsub:data_owner}
A data owner is an individual person associated with an IoT device such as cell phone, smart car, and body sensor. The device has the owner’s location, health associated details, and financial transaction details etc. The data is then collected by the server (curator). In our system model, we consider a set of data owners which is denoted as $\textbf{O} = \{O_{1}, O_{2}, O_{3}...O_{n} \}$ where $n$ is the number of data owners.
\subsubsection{Data User}
\label{subsub:data_user}
A data user is a third-party organization, company or governmental agency which needs exploratory analysis of the data collected by the curator. We assume a set of data users denoted as $\textbf{U} = \{U_{1}, U_{2}, U_{3}...U_{k}\}$ where $k$ is the number of data users.
\subsubsection{Query}
\label{subsusb:query}
A query represents a statistical query such as Cunt, Average, Maximum, and Minimum which is denoted as $q$. The data users from the set \textbf{U} send queries to the data curator \textbf{C} which are then evaluated by the data curator over the actual dataset. Afterwards, a random noise is inserted into the query response to perturb it before sharing it with the data user which is denoted as $q_{'}$. 
\subsubsection{Blockchain Network}
\label{subsub:block-net}
The blockchain network is based on Hyperledger fabric as shown in Fig. \ref{fig:system_model}. Data curator and data user act as complete organizations with its own databases. Furthermore, each of these represents a Hyperledger fabric node in the network. Therefore, the data sharing event is recorded as a transaction on the blockchain ledger. The contents of the transaction include query type, and differential privacy budget $\epsilon$ which is utilized in generating a perturbed query response.  
\subsubsection{Query and Verification by Data Owner}
\label{subsub:qu_ver_by_data_owner}
To make the data sharing event transparent and avoid the low-level threat of privacy breach of data curator, each data owner sends query to the Hyperledger fabric network. Furthermore, for the query to be evaluated on the blockchain ledger, query transaction of Hyperledger fabric is adopted \cite{hyper}. Afterwards, the response is returned with the query type, and the privacy budget utilized to the concerned data owner \textbf{O}.  
\subsection{Threat Model of OPU-TF-IoT}
\label{subsec:tht_model}
Two types of adversaries exist in the proposed system model which are (i) the centralized curator, and (ii) third parties (companies, organizations, advertisement agencies etc.). Furthermore, the curator can act as honest-but-curious adversary. To this end, data owners trust on the curator that it will ensure the privacy protection of sensitive information while sharing the aggregated results with the third parties. However, due to lack of transparency in the operations of the curator in the traditional approaches, it can share the data with loose privacy preservation, i.e., by using a large value of $\epsilon$ for its own benefit. Therefore, in this work, the threat from the data curator is regarded as average level threat. \\
On the other hand, third parties cause serious threats to the privacy of data owners because of their strong background knowledge. Consequently, despite of using a suitable privacy budget for perturbation of query response, it is more likely that an individual can be exposed through linking or inference privacy attack. In a linking privacy attack, an adversary uses the perturbed data and link it with the background knowledge to get the actual data of a data owner. Similarly, in inference privacy attack, an adversary tries to predict the actual data of a data owner based on mathematical or statistical techniques such as average, median etc. Furthermore, in real-world scenarios, the individual data records may be correlated which further increases the risk of privacy breach. Therefore, care must be taken to avoid the privacy breach by using a suitable privacy budget agreed between the curator and data owner. Moreover, because the data records in real-world scenarios are often correlated therefore, the curator should use the composition theorem, i.e., \textit{Theorem 2} given in Section \ref{subsubsec:dp} to keep an eye on the maximum privacy budget. \\ 
In both cases, the privacy breach can result in the exposure of sensitive information of data owners such as life style, shopping activities, location visited, choice, financial status, social relationships, and political beliefs.  

%%%% Algorithm 1
\begin{algorithm}[tp!]
\small
\let\oldnl\nl% Store \nl in \oldnl
\newcommand{\nonl}{\renewcommand{\nl}{\let\nl\oldnl}}
\caption{Heuristic search in OPU-TF-IoT (note: step 1 and 20 of our proposed algorithm have been taken from \cite{trackbudget})}
\label{alg:heu_search}
\SetAlgoLined
\DontPrintSemicolon
\nonl

\textbf{Input:} privacy budget $\epsilon$, data table $T_{n, m}$, set of data users $\textbf{U} = \{U_{1}, U_{2}, U_{3}...U_{k}\}$, required accuracy $A^{i}_{req}$ by data user $U_{i}$, query function $F$, query type $N$, query $q_{i}$\;
\nonl

\textbf{Output:} perturbed query response $q^{'}_{i}$ with suitable privacy budget $\epsilon_{sut}$\;
\nonl

\textbf{Initialization:} iteration \textit{i = 1}, total privacy budget = $\epsilon_{t}$, default privacy budget $0 < \epsilon_{def} < \epsilon_{t}$, \textit{x} is a random variable, \textit{noise} = 0, mean $\mu$ = 0, sensitivity $\bigtriangleup$f = 1, Laplace scale $\lambda = \frac{{\bigtriangleup}f}{\epsilon}$, $q_{i} = [A^{i}_{req}, F, N]$, tolerance factor = $\tau$, decrement factor $0 < \eta < 1$\;

\While (\tcp*[h]{check the budget availability and data users}){\textit{$i \leq k \land \epsilon_{t} \leq \epsilon_{def}$}} {
     parse the parameters $A^{i}_{req}$, $F$, $N$ from $q_{i}$ of $U_{i}$ \;
     classify the $q_{i}$ based on the value of $N$ \tcp*[h]{\ref{par:query_type_N}} \;
     \textbf{Call} QueryFunction($q_{i}, \epsilon_{def}$)\;
     get $A^{i}_{act}$ \tcp*[h]{using equation \ref{eq:accur}}\;	
     
     \If (\tcp*[h]{from inequality \ref{ineq:tolerance}}){$|A^{i}_{act}-A^{i}_{req}| \leq \tau$} {   
				$\epsilon_{t} \leftarrow \epsilon_{t} - \epsilon_{def}$ \tcp*[h]{decrement $\epsilon_{t}$}\; 				
				$q^{'}_{i} \leftarrow q_{i}$\;
	 }% end of if 	 
	 
	 \ElseIf {($|A^{i}_{act}-A^{i}_{req}| \not\leq \tau) \land (A^{i}_{act} < A^{i}_{req}$)}{							
				needs an alternative plan \;
				skip the query $q_{i}$ \;				
     } %end of elseif 
	 
	 \Else {
				$\epsilon_{sut} = \epsilon_{def}$\;				
				\While {$|A^{i}_{act}-A^{i}_{req}| \not\leq \tau$}{
				\Indp 								
				$\epsilon_{sut} = \epsilon_{sut} - \eta$ \tcp*[h]{decrement by $\eta$}\;				 
				} % end this while 
    \textbf{Call} QueryFunction($q_{i}, \epsilon_{sut}$)\;	
    $\epsilon_{t} \leftarrow \epsilon_{t} - \epsilon_{sut}$ \tcp*[h]{decrement $\epsilon_{t}$}\;	
	\KwRet $q^{'}_{i}$\;
	 }%end of else		

%\nonl: No line Numbre 
%\Indp Indentation
	\nonl

	$\textbf{FUNCTION}\rightarrow QueryFunction(q_{i}, \epsilon)$\;  
	\Indp 
		evaluate query $q_{i}$ on the data table $T_{n, m}$\;
		\textbf{Call} LaplacianFunction($q_{i}, \epsilon$)\;
		 $q^{'}_{i} \leftarrow q_{i} + noise$ \tcp*[h]{add noise into $q_{i}$}\;
		\KwRet {$q^{'}_{i}$}\;
		\Indm
		\nonl

		$\textbf{FUNCTION}\rightarrow LaplacianFunction(q_{i}, \epsilon)$\;  
		\Indp 
			generate noise using \textit{$f(x;\mu,\frac{{\bigtriangleup}f}{\textit{$\epsilon$}})$}\tcp*[h]{equation \ref{eq:lap_dist}}\; 
			\KwRet {$noise$}\tcp*[h]{Laplacian random noise}\;
		\Indm
$i \leftarrow i + 1$
} %Outer while
\KwRet $\{q^{'}_{1}, q^{'}_{2}, q^{'}_{3}… q^{'}_{k}\}$ \tcp*[h]{set of queries responses with adjusted privacy budget}\;
\end{algorithm}
%%% end algorihtm 1

\subsection{Framework Design}
\label{subsec:fram_design}
The framework design is shown in Fig. \ref{fig:fram_design}. The data curator \textbf{C} collects the data from the set of data owners \textbf{O}. At the same time, the privacy preservation level, i.e., privacy budget for each data owner is also collected which results in a set of privacy budget values denoted as $\bm{\epsilon} = \{\epsilon_{1}, \epsilon_{2}, \epsilon_{3} ... \epsilon_{n}\}$. Each data owner wants to decrease the privacy leakage by adopting a small value of $\epsilon$. Afterwards, the data is recorded in the form of a table with columns and rows which is denoted as \textbf{$T_{n,m}$} whereas $n$ represents the number of rows and $m$ represents the number of columns. More specifically, the $n^{th}$ row represents the record of the $n^{th}$ data owner in the dataset. Similarly, the $m^{th}$ column represents the $m^{th}$ attribute of the record. For instance, in a medical dataset, each row represents record of a patient while each column represents a specific disease. For simplicity, we assume that the data curator selects the minimum privacy budget from the list $\bm{\epsilon}$ as the total privacy budget $\epsilon_{t}$ which will satisfy the privacy requirements of all the data owners. Furthermore, the rows of $T_{n,m}$ are considered as correlated which means if an individual from the table is isolated by the adversary then the risk of privacy breach of other related records increases \cite{gambasedpriv_utility}. \\
The data user sends the query $q_{i}$ which consists of three parameters denoted as $[A^{i}_{req}, F, N]$ where $A^{i}_{req}$ is the required accuracy, $F$ is the query function and $N$ is the query type defined as follows:
\paragraph{Accuracy $A^{i}_{req}$} it denotes the required accuracy of the query response set by the data user $U_{i}$. For accuracy, we need to define the relative error in the query response. According to \cite{relterror}, the relative error $r^{i}_{err}$ of the $i^{th}$ query response can be defined as follows:
\begin{align}
r^{i}_{err} = \frac{|\Upsilon_{i} - \Upsilon^{'}_{i}|}{\Upsilon_{i}} && \text{(as stated in \cite{relterror})} \label{eq:rel_err}
\end{align}
where $\Upsilon_{i}$ and $\Upsilon^{'}_{i}$ are the numerical values of $q_{i}$ and $q^{'}_{i}$, respectively.\\
Based on $r^{i}_{err}$ and the required accuracy $A^{i}_{req}$ of data user $U_{i}$, the curator \textbf{C} calculates the the actual accuracy $A^{i}_{act}$ of the query response as follows:
\begin{align}
A^{i}_{act} = 1 – r^{i}_{err} %&& \text{(as stated in \cite{relterror})} 
\label{eq:accur}
\end{align} 
where $0 \leq r^{i}_{err} \leq 1$ and hence, $0 \leq A^{i}_{act} \leq 1$. As a result, 0 means minimum accuracy and 1 means maximum accuracy.  \\
Consequently, to satisfy the data user $U_{i}$, $A^{i}_{act} \geq A^{i}_{req}$. Furthermore, if $A^{i}_{act} < A^{i}_{req}$ then an alternative plan is needed which should be agreed by the curator and the data user. For instance, the data user can compensate the accuracy, or the curator can generate more accurate query response with the consent of data owners to satisfy the data user.   
\paragraph{Query Function $F$} it denotes the query function which is given as $F \in \{Count, Average, Maximum, Minimum\}$. Each element of F represents a category of statistical query. Therefore, the curator uses the value of F to evaluate the associated query on the $T_{n,m}$. 

\begin{figure*}[!hpt]
\begin{center}
\includegraphics[width=13.5cm]{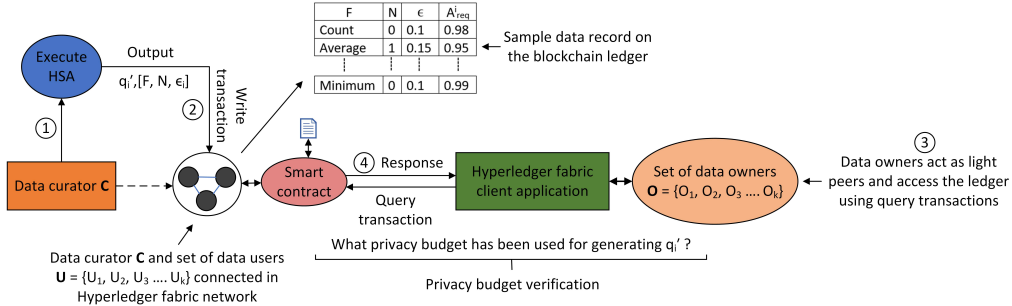}
\caption{Overview of the privacy budget verification mechanism in the proposed framework (OPU-TF-IoT).}
\label{fig:verif_design}
\end{center}
\end{figure*}

\paragraph{Query Type $N$} 
\label{par:query_type_N}
it denotes the query type which is defined as $N \in \{0, 1\}$. Here, $N = 0$ represents that the data user wants the query to be evaluated on the whole dataset whereas N = 1 represents that data user is only interested in a part of the dataset. For instance, if the patients medical records throughout the USA is considered then the two types of queries are given below.  
\begin{enumerate}
\item For N = 0, $q_{i}$ is how many patients throughout the USA have suffered from disease $x$?  
\item For N = 1, $q_{i}$ is how many patients suffered from a disease $x$ in the New York region? 
\end{enumerate}     

Therefore, by considering the parameters $[A^{i}_{req}, F, N]$ of $q_{i}$, the curator then uses the proposed HSA to decide a suitable $\epsilon$ and generate $q^{'}_{i}$ as shown in the Fig. \ref{fig:fram_design}. The details of HSA are given in the next section.\\
Based on \textit{Definition 1} and \textit{Theorem 2}, we define the guarantee of privacy preservation against the adversaries discussed in section \ref{subsec:tht_model} as following: \\
\textbf{Definition 2:} If $\textbf{M} = \{M_{1}, M_{2}, M_{3} … M_{k}\}$ represents the set of mechanisms for calculating the responses to the queries sent by the set of data users \textbf{U} on the data table $T_{n,m}$ with the condition that each mechanism $M_{i} \in \textbf{M}$ satisfies $\epsilon_{i}$-differential privacy, then the \textbf{M} satisfies $(\sum_{i=1}^{k} \epsilon_{i})$-differential privacy.\\
Apart from this, the utilization of data is defined in terms of accuracy $A_{act}$ of the query responses. Here, we use $A_{act}$ without the superscript $i$ to denote accuracy in general and not for the $q_{i}$. Furthermore, due to the random noise addition, it is very difficult to get a smooth value of the actual accuracy $A_{act}$ so that $A_{act} \geq A_{req}$. Hence, a tolerance coefficient $\tau$ is introduced such that $0 \leq \tau \leq 1$, and it is defined as the fraction by which a data user can tolerate the accuracy. Consequently, we define the utilization of the data as following:\\
\textbf{Definition 3:} For given values of $A_{act}$, $A_{req}$, and $\tau$, the utilization of the data is satisfactory if the following holds:
\begin{align}
|A_{act}-A_{req}| \leq \tau && %\text{(as stated in \cite{relterror})} 
\label{ineq:tolerance}
\end{align}     
Furthermore, if the inequality in \ref{ineq:tolerance} does not hold then an alternate plan is needed as discussed earlier.    
\subsubsection{Heuristic Search Algorithm (HSA)}
\label{subsub:hsa}
The data curator executes the proposed algorithm \ref{alg:heu_search} to adopt a suitable value of $\epsilon$ denoted as $\epsilon_{sut}$ for generating a perturbed query response $q^{'}_{i}$. It is to be noted that two steps for managing the privacy budget, i.e., step 1 ($\epsilon_{t} \leq \epsilon_{def}$) and step 20 ($\epsilon_{t} \leftarrow \epsilon_{t} - \epsilon_{sut}$) of our proposed algorithm have been taken from \cite{trackbudget}. The reason is that we have further improved the previous algorithm proposed in \cite{trackbudget}. Furthermore, $\epsilon_{sut}$ is the minimum value of the privacy budget which satisfies the accuracy requirements of the $q_{i}$ by a $U_{i}$. Based on the parameters $[A^{i}_{req}, F, N]$, the curator first uses a default privacy budget $\epsilon_{def}$ which is selected randomly according to $0 < \epsilon_{def} < \epsilon_{t}$ to generate $q^{'}_{i}$. Later in this work, we will propose an algorithm for the curator to choose the default privacy budget $\epsilon_{def}$ in order to further optimize the privacy-utility trade-off. Afterwards, the curator calculates the accuracy $A^{i}_{act}$ by using equation \ref{eq:accur}. To minimize the effect of randomness of Laplacian noise, the curator generates a vector of noise values of length 1000 by using the same $\epsilon$. Similarly, the associated $A^{i}_{act}$ is calculated for each noise value by using equation \ref{eq:accur}. Subsequently, the average $A^{i}_{act} = \frac{\sum_{j=1}^{1000}A^{j}_{act}}{1000}$ is then compared with the $A^{i}_{req}$ value of $q_{i}$. According to inequality \ref{ineq:tolerance}, the data curator can take three types of decisions which are given below.
\begin{enumerate}
\item If $|A^{i}_{act}-A^{i}_{req}| \leq \tau$ then the utilization is satisfactory and the $q^{'}_{i}$ is returned to the data user. 
\item If $|A^{i}_{act}-A^{i}_{req}| \not\leq \tau$, and $A^{i}_{act} < A^{i}_{req}$, then the data user is not satisfied, and an alternative plan is needed. 
\item If $|A^{i}_{act}-A^{i}_{req}| \not\leq \tau$, and $A^{i}_{act} > A^{i}_{req}$, then the data curator needs to adjust the $\epsilon$ in order to avoid the waste of privacy budget. 
\end{enumerate}
In case 3 above, we introduce a decrement factor denoted as $\eta$ such that $0 < \eta < 1$ which is used to decrement the default $\epsilon_{def}$ until the condition $|A^{i}_{act}-A^{i}_{req}| \leq \tau$ is satisfied. The detailed steps of the three cases are given in lines 6, 10, and 16 of algorithm \ref{alg:heu_search}, respectively. Finally, the perturbed query response set $\{q^{'}_{1}, q^{'}_{2}, q^{'}_{3}… q^{'}_{k}\}$ is retuned as the output of the algorithm \ref{alg:heu_search}. The output consists of all those queries which satisfy the accuracy requirements $\{A^{1}_{req}, A^{2}_{req}, A^{3}_{req}… A^{k}_{req}\}$ of the data users. In addition, the queries which fail to satisfy the accuracy requirements are skipped as shown in lines 10-13 of the algorithm \ref{alg:heu_search}. Furthermore, to maximize the number of satisfied users $U_{i} \in \textbf{U}$ by minimizing the number of skipped queries, the data curator selects suitable values of $\epsilon_{def}$ and $\eta$. The selection of $\epsilon_{def}$ and $\eta$ by the curator is presented in the following section.\\    
The novelty of the algorithm \ref{alg:heu_search} is that it finds a suitable privacy budget $\epsilon_{sut}$ by reducing the default privacy budget $\epsilon_{def}$ as shown in lines 16-20 of algorithm \ref{alg:heu_search}. As a result, the waste of privacy budget is avoided whereas both data owners and data users are satisfied. In the following, we discuss the selection of $\epsilon_{def}$ and $\eta$ in detail.
\subsubsection{Selection of $\epsilon_{def}$ and $\eta$}
\label{subsec:sel_of_ep_and_eta} 
The values of $\epsilon_{def}$ and $\eta$ impact the number of satisfied data users. Furthermore, the curator tries to keep the number of satisfied data users above a threshold $\rho$ such that $\rho \leq k$ which defines the minimum number of satisfied data users from the set \textbf{U}. For instance, if the curator starts with a smaller value of $\epsilon_{def}$ in the range $0 < \epsilon_{def} < \epsilon_{t}$ in algorithm \ref{alg:heu_search} then the data users with high accuracy requirements may not be satisfied due to less accurate query responses (line 6 of algorithm \ref{alg:heu_search}). The reason is that a smaller value of $\epsilon_{def}$ leads to high noise addition into the query response. On the other hand, a relative high value of $\epsilon_{def}$ in algorithm \ref{alg:heu_search} will satisfy most of the data users because of the less noise addition into the query responses. However, using a high value of $\epsilon_{def}$ will lead to quick exhaustion of the total privacy budget $\epsilon_{t}$ as shown in lines 7 and 20 of algorithm \ref{alg:heu_search}. The reason is that a relatively high privacy budget is utilized to generate individual query response which accumulates to a high value according to \textit{Theorem 2}. \\
Apart from the $\epsilon_{def}$, the curator also selects a suitable value of $\eta$ to gradually decrease the $\epsilon_{def}$ as shown in lines 15-18 of the algorithm \ref{alg:heu_search}. A smaller value of $\eta$ will decrease the $\epsilon_{def}$ in a more granular manner to find a best fit $\epsilon_{sut}$. Similarly, a relative high value of $\eta$ may not find $\epsilon_{sut}$ to satisfy the condition given in line 16 of algorithm \ref{alg:heu_search}. Consequently, the associated $q_{i}$ will be skipped which is not desired. Therefore, the curator uses algorithm \ref{alg:sel_of_ep_and_eta} to find a suitable $\epsilon_{def}$ at the beginning and a suitable $\eta$ which is used to find a best fit $\epsilon_{sut}$. In algorithm \ref{alg:sel_of_ep_and_eta}, the curator first picks the values of  $\epsilon_{def}$ and $\eta$ from the current execution of algorithm \ref{alg:heu_search}. Afterwards, it checks the number of satisfied data users against the threshold $\rho$. Consequently, it enables the curator to select best values of $\epsilon_{def}$ and $\eta$ which not only decrease the privacy budget utilization but also satisfy the accuracy requirements of all the data users. Moreover, to enable the verification of the utilization of privacy budget in OPU-TF-IoT, the following section discusses the proposed verification mechanism. 

%%%% Algorithm 2
\begin{algorithm}[tp!]
\small
\let\oldnl\nl% Store \nl in \oldnl
\newcommand{\nonl}{\renewcommand{\nl}{\let\nl\oldnl}}
\caption{Selection of $\epsilon_{def}$ and $\eta$ in OPU-TF-IoT}
\label{alg:sel_of_ep_and_eta}
\SetAlgoLined
\DontPrintSemicolon
\nonl
\textbf{Repeat:}\;

get the values of $\epsilon_{def}$ and $\eta$ from the current execution of algorithm \ref{alg:heu_search} \;
get no of satisfied data users from the current execution of algorithm \ref{alg:heu_search} \;
\If {$no\_of\_satisfied\_data\_users < \rho$} {   
	increase the current $\epsilon_{def}$ and decrease the previous $\eta$ for the next execution of algorithm \ref{alg:heu_search}\; 				
	}% end of if 	 
\Else {
	   continue\;								 
	  } %  	
\end{algorithm}
%%% end algorihtm 2
 
\subsubsection{Privacy Budget Verification Mechanism}
\label{subsecb:veri_mech}
Blockchain-based verification mechanism is shown in Fig. \ref{fig:verif_design}. The verification mechanism uses smart contract, write transactions, query transactions, and client application of the Hyperledger fabric which are defined as following \cite{hyper}.
\paragraph*{Smart contract} the smart contract of Hyperledger fabric is known as chaincode. An instance of the smart contract is installed on each peer or node of the Hyperledger fabric network. The smart contract defines the functions which operates on the blockchain ledger such as write, read, query etc. 
\paragraph*{Write transaction} it invokes the smart contract function which alters the records on the ledger. Therefore, a write transaction changes the state of the ledger. 
\paragraph*{Query transaction} it invokes the function of smart contract which evaluate the result of a query on the ledger. Furthermore, it does not change the state of the ledger, i.e., the query is evaluated and returned to the requester. 

\begin{figure*}[!htp]
\centering
\begin{subfigure}[t]{0.49\textwidth}
\centering
\includegraphics[height = 4cm]{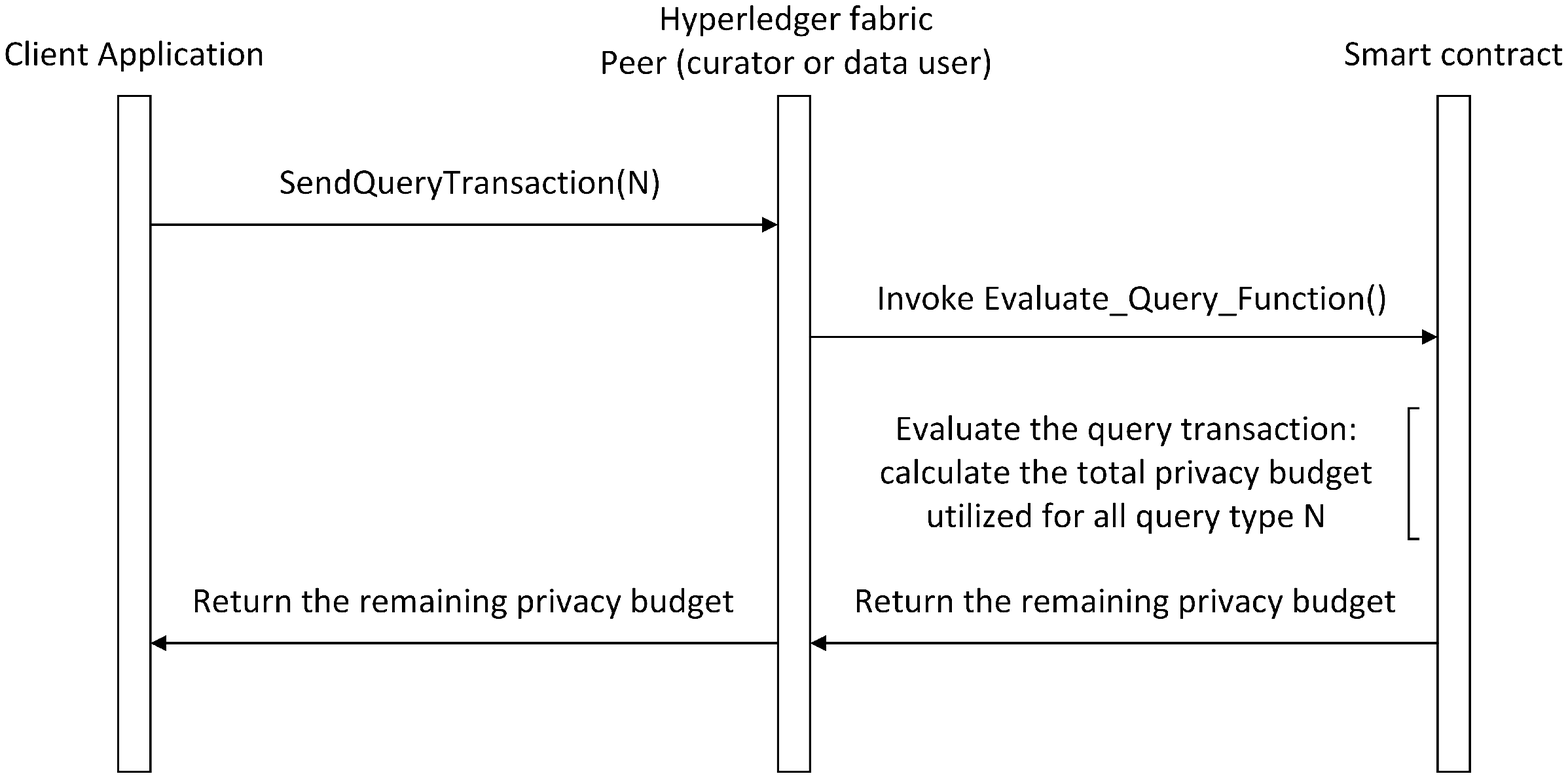}
\caption{ }
\label{subfig:flow_diagram}
\end{subfigure} %\hspace{8mm}
\begin{subfigure}[t]{0.49\textwidth}
\centering
\includegraphics[height = 4cm, width=0.99\textwidth]{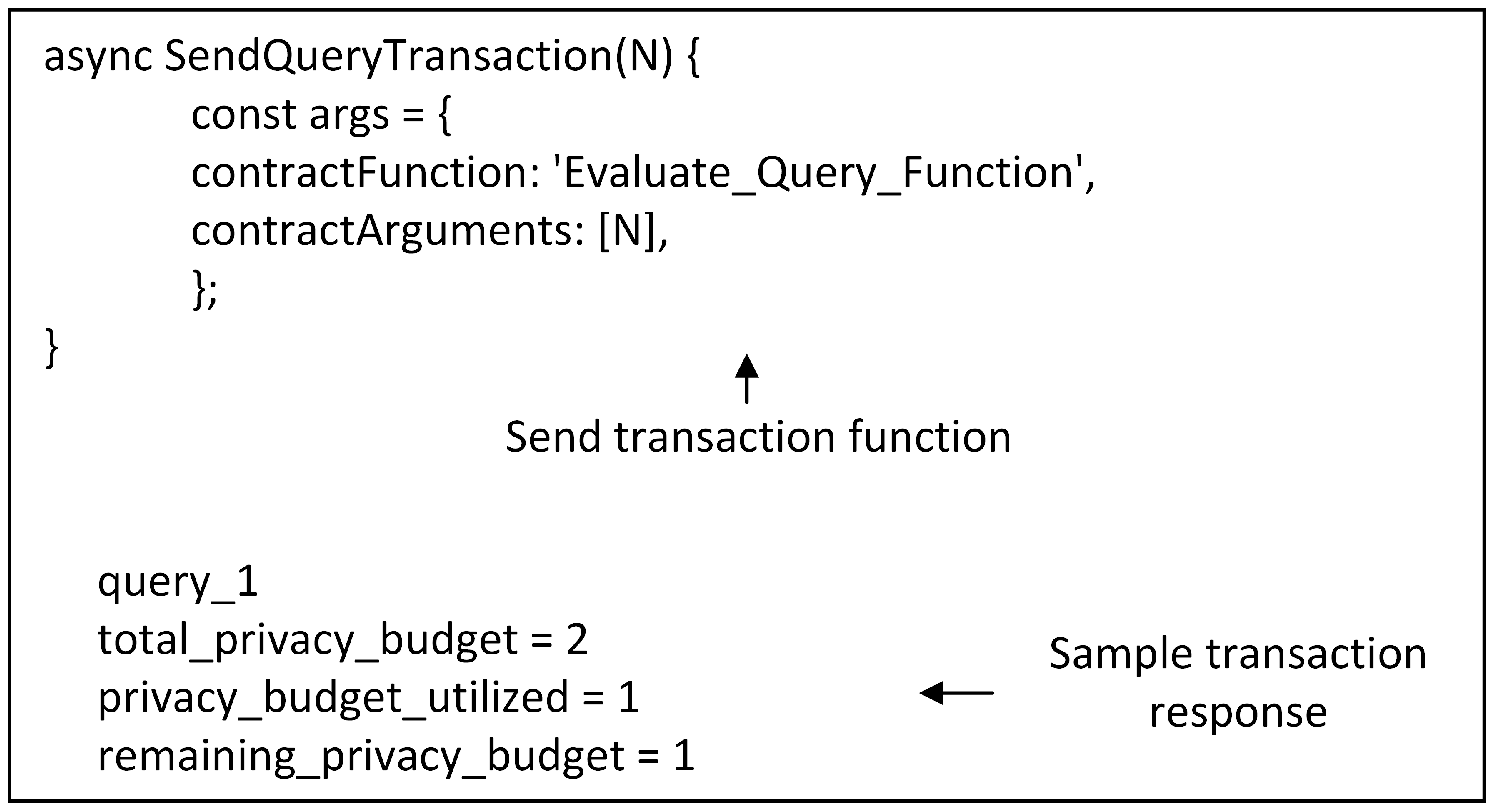}
\caption{ }
\label{subfig:tran_response}
\end{subfigure}
\caption{(a) Working flow of the proposed privacy budget verification mechanism, (b) illustration of transaction and transaction response.}
\label{fig:flow_dia+tran_resp}
\end{figure*}

\paragraph*{Client application} it is used to access the ledger through query transactions in Hyperledger fabric network. In the proposed scenario, the data owners act as client applications. Moreover, the data owners \textbf{O} are light peers of the blockchain network which means that it can only access the ledger state but cannot modify it. On the other hand, the data curator \textbf{C} and the set of data users \textbf{U} are full peers which means they have full rights of modifying and setting the policies for the rest of the network. 
\paragraph*{Consensus} in the proposed work, the deterministic consensus mechanism of Hyperledger fabric is adopted in which specified peers called orderer peers performs the consensus process \cite{hyper}. In the proposed scenario, data curator and data users are responsible for carrying out the consensus, validation of transactions, and configuration of the smart contract policies of the network. \\
The parameters $[F, N, \epsilon_{i}, A^{i}_{req}]$ along with the $q^{'}_{i}$ are recorded on the Hyperledger fabric ledger as shown in Fig. \ref{fig:verif_design}. Therefore, the record of the privacy budget utilization for each successful query is maintained. The client applications then send query transactions which are evaluated on the ledger and returned to the requestors. The working flow, associated functions of the smart contract, and sample response are shown in Fig. \ref{fig:flow_dia+tran_resp}. In Fig.  \ref{fig:flow_dia+tran_resp}(a), client application sends transaction with the parameter $N$ using the SendQueryTransaction() which invokes the associated Evaluate\_Query\_Function() of the smart contract. Subsequently, the query is evaluated on the ledger according to the value of $N$ such that if $N = 0$ then according to \textit{Theorem 2}, the privacy budget utilized is equal to $\sum_{i=1}^{z} \epsilon_{i}$ where $\epsilon_{i}$ is the fraction of privacy budget used for generating $q^{'}_{i}$, and $z$ is the number of all queries for which $N = 0$. Similarly, if $N = 1$ then the privacy budget utilized is equal to $\sum_{i=1}^{z} \epsilon_{i}$ where $z$ is the number of queries for which $N= 1$. Furthermore, the sample response consists of the total privacy budget, utilized privacy budget, and the remaining privacy budget as shown in Fig. \ref{fig:flow_dia+tran_resp}(b). In this way, the data owners can verify and track the privacy budget utilization in each query. As a result, it satisfies the data owners regarding the use of their private data. \\
Apart from the verification and tracking of the privacy budget, utilizing the previous response of a repeated query can also save the accumulated privacy budget \cite{trackbudget}. Therefore, in OPU-TF-IoT, the curator searches the recorded query responses before utilizing new privacy budget using algorithm \ref{alg:utiliz_prev_budget}. Consequently, if the required accuracy $A^{i}_{req}$, query function $F$, and query type $N$ of the incoming query $q_{i}$ match with any of the record on blockchain ledger then it is returned to the data user without utilizing a new privacy budget as shown in lines 1-2 of the algorithm \ref{alg:utiliz_prev_budget}. In this way, the utilization of privacy budget is further decreased. In the following sections, we present the time complexity, performance evaluation and comparison of the proposed work with the state-of-the-art works.  

%%%% Algorithm 3
\begin{algorithm}[tp!]
\small
\let\oldnl\nl% Store \nl in \oldnl
\newcommand{\nonl}{\renewcommand{\nl}{\let\nl\oldnl}}
\caption{Utilization of the previous privacy budget in OPU-TF-IoT}
\label{alg:utiliz_prev_budget}
\SetAlgoLined
\DontPrintSemicolon
\nonl
\textbf{Repeat:}\;
\If {$[A^{i}_{req}, F, N] == any\_record\_on\_the\_ledger$} {   
	 $q^{'}_{i} \leftarrow record\_on\_blockchain\_ledger$;\	 			
	}% end of if 	 
\Else {
	   continue with the execution of algorithm \ref{alg:heu_search}\;								 
	  } %  	
\end{algorithm}
%%% end algorihtm 3

\subsubsection{Time Complexity of the Proposed Algorithms}
\label{subsub:tim_com}
In this section, we discuss the time complexity of the proposed algorithms. Furthermore, as algorithm \ref{alg:heu_search} performs the main implementation task of the proposed OPU-TF-IoT so, we only evaluate the time complexity of algorithm \ref{alg:heu_search}. Algorithm \ref{alg:heu_search} consists of two \textit{while} loops which are the outer and inner \textit{while} loops given in line 5 and 16, respectively. The outer \textit{while} loop executes according to the size of \textbf{U} whereas the inner \textit{while} loop only executes when the $\epsilon_{def}$ need to be adjusted. In real-world scenarios, the $\epsilon_{def}$ is not necessarily adjusted for all data users. \\
Similarly, for typical values of $\epsilon_{def}$ in the range [0.1, 1], the inner \textit{while} loop takes around 1000 steps to reduce $\epsilon_{def} = 1$ by 50\%. Therefore, the worst-case time complexity of algorithm \ref{alg:heu_search} is calculated as $|\textbf{U}|$*1000 where $|\textbf{U}|$ denotes the size of \textbf{U}. Consequently, the time complexity O($|\textbf{U}|$) = 1000$|\textbf{U}|$. As a result, a cloud server can easily execute the proposed algorithm \ref{alg:heu_search}.  

\begin{figure}[bp]
\begin{center}
\includegraphics[width=\linewidth]{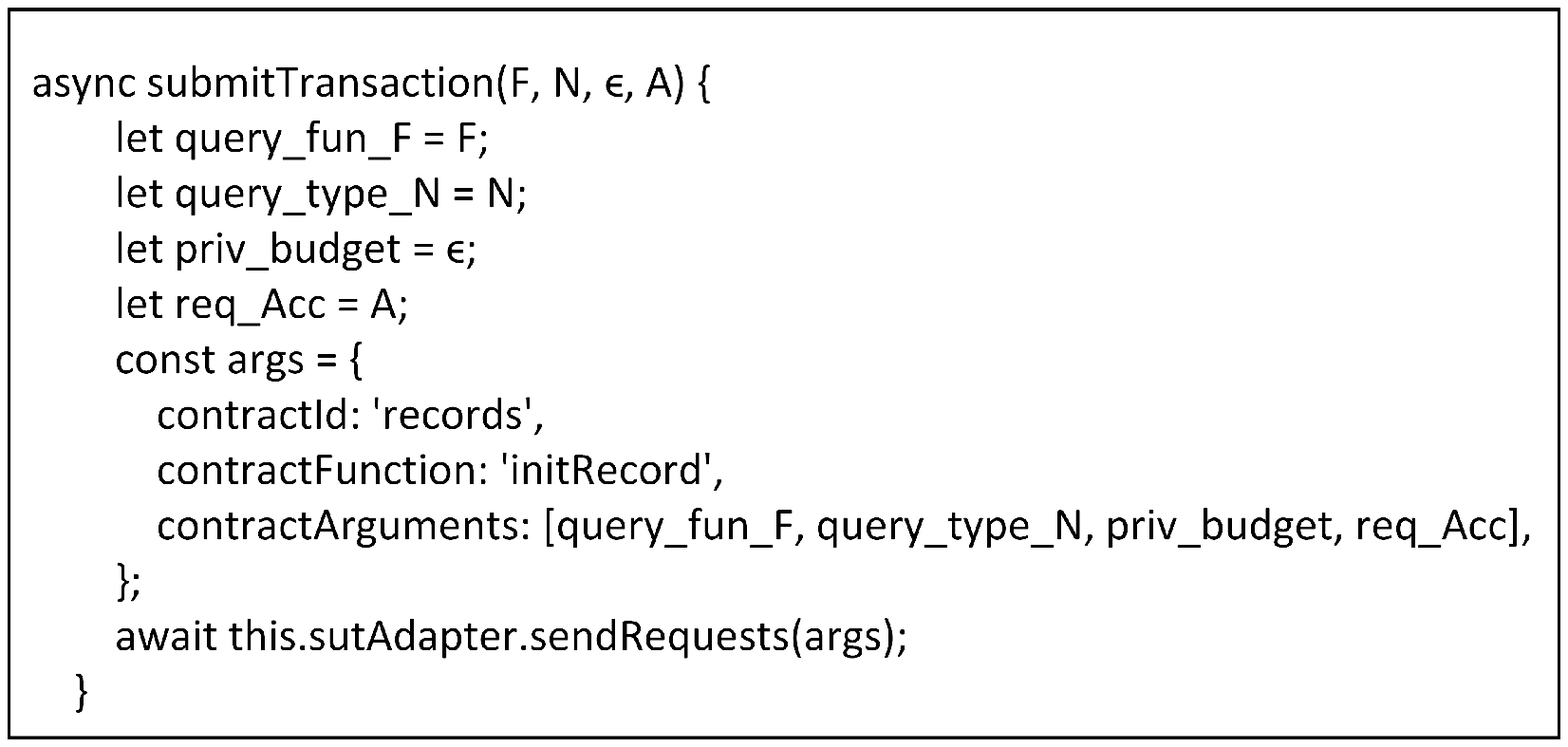}
\caption{Write transaction initialization in OPU-TF-IoT.}
\label{fig:write_tran_init}
\end{center}
\end{figure}

\section{Performance Evaluation}
\label{sec:per_eval}
In this section, we present the performance evaluation of the proposed work and its comparison with state-of-the-art works. The state-of-the-art works include the standard differential privacy model (standard DP) presented in \cite{dpint2006} and a blockchain-based approach for saving and tracking differential-privacy cost (BST-DP) \cite{trackbudget}. Furthermore, the performance is evaluated over three parameters which are (1) optimized privacy-utility trade-off (2) verification of privacy budget utilization, and (3) impact of $\tau$ and $\eta$ on the performance of OPU-TF-IoT. Firstly, we discuss the datasets and simulation setup then the results and discussion are presented.

%%% trade-off
\begin{figure*}[!htb]
\centering
\begin{subfigure}[t]{0.48\linewidth}
\centering
\includegraphics[width=\linewidth]{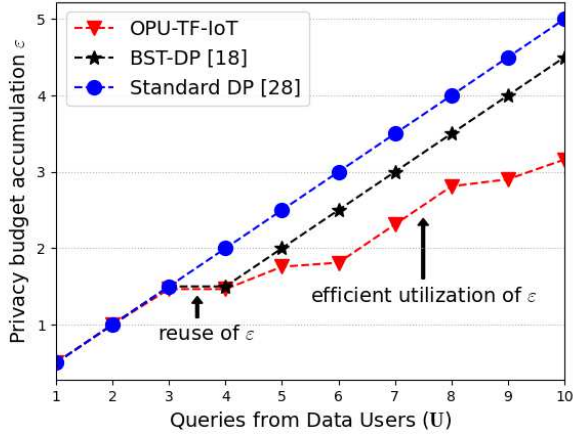}
\caption{Count queries}
\label{count_query}
\end{subfigure}
\begin{subfigure}[t]{0.48\linewidth}
\centering
\includegraphics[width=\linewidth]{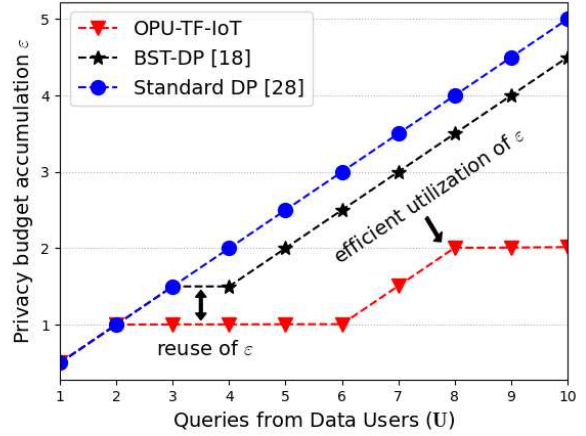}
\caption{Average queries}
\label{avg_queries}
\end{subfigure}
\centering
\begin{subfigure}[t]{0.48\linewidth}
\centering
\includegraphics[width=\linewidth]{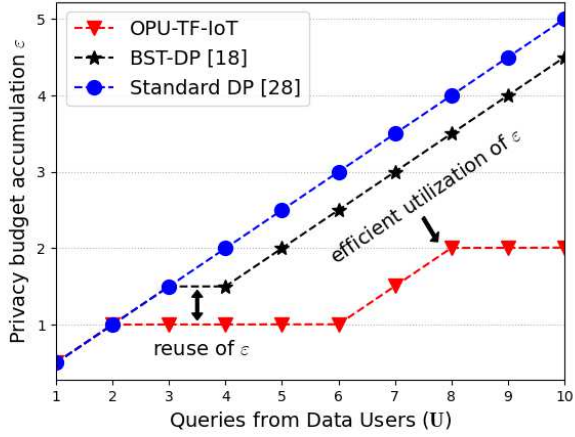}
\caption{Maximum queries}
\label{max_queries}
\end{subfigure}
\centering
\begin{subfigure}[t]{0.48\linewidth}
\centering
\includegraphics[width=\linewidth]{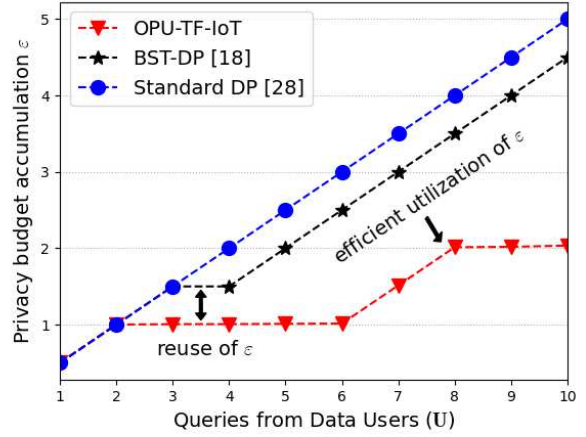}
\caption{Minimum queries}
\label{min_queries}
\end{subfigure}
\caption{Evaluation and comparison of privacy-utility trade-off for OPU-TF-IoT, BST-DP \cite{trackbudget}, and Standard DP \cite{dpint2006} with $\epsilon_{def} = 0.5$, $\tau = 0.02$, and $\eta = 0.0005$.}
\label{fig:eval_and_com_of_priv_util_trad_off}
\end{figure*}

\begin{table*}[h]
\centering
\caption{Comparison of total privacy budget utilization for OPU-TF-IoT, standard DP \cite{dpint2006}, and BST-DP \cite{trackbudget} with $\tau = 0.02$ and $\eta = 0.0005$ where Count, Avg, Max, and Min represent Count, Average, Maximum, and Minimum queries, respectively.}
\begin{center}
\begin{tabularx}{\linewidth}{|Y|Y|Y|Y|Y|Y|Y|Y|Y|Y|Y|Y|Y|Y|} 
\hline
\multirow{2}{*}{$\epsilon_{def}$} & \multicolumn{4}{c|}{\textbf{Total privacy budget in OPU-TF-IoT}} & \multicolumn{4}{c|}{\textbf{Total privacy budget in BST-DP \cite{trackbudget}}} & \multicolumn{4}{c|}{\textbf{Total privacy budget in Standard DP \cite{dpint2006}}} \\
\cline{2-13}
 & Count & Avg & Max & Min & Count & Avg & Max & Min & Count & Avg & Max & Min \\ 
\hline
 0.1 & 0.45	& 0.41 & 0.4 & 0.43 & 0.5 & 0.89 & 0.89 & 0.89 & 0.5 & 0.99 & 0.99 & 0.99 \\
\hline
 0.2 & 1.32 & 0.81 & 0.80 & 0.83 & 1.59 & 1.79 & 1.79 & 1.79 & 1.79 & 1.99 & 1.99 & 1.99 \\
\hline
 0.3 & 2.17 & 1.21 & 1.20 & 1.23 & 2.69 & 2.69 & 2.69 & 2.69 & 2.99 & 2.99 & 2.99 & 2.99 \\
\hline
 0.4 & 2.65 & 1.61 & 1.60 & 1.63 & 3.59 & 3.59 & 3.59 & 3.59 & 3.99 & 3.99 & 3.99 & 3.99 \\
\hline
 0.5 & 3.12 & 2.01 & 2.0 & 2.03 & 4.5 & 4.5 & 4.5 & 4.5 & 5 & 5 & 5 & 5 \\
\hline
 0.6 & 3.48 & 2.41 & 2.40 & 2.43 & 5.39 & 5.39 & 5.39 & 5.39 & 5.99 & 5.99 & 5.99 & 5.99 \\
\hline
 0.7 & 3.96 & 2.81 & 2.8 & 2.83 & 6.3 & 6.3 & 6.3 & 6.3 & 7 & 7 & 7 & 7 \\
\hline
 0.8 & 4.38 & 3.21 & 3.2 & 3.23 & 7.19 & 7.19 & 7.19 & 7.19 & 7.99 & 7.99 & 7.99 & 7.99 \\
\hline
 0.9 & 4.79 & 3.61 & 3.6 & 3.63 & 8.10 & 8.1 & 8.1 & 8.1 & 9 & 9 & 9 & 9 \\
\hline
 1 & 5.18 & 4.01 & 4 & 4.03 & 9 & 9 & 9 & 9 & 10 & 10 & 10 & 10 \\
\hline
\end{tabularx}
\end{center}
\label{tab:com_of_tot_budg}
\end{table*}

\begin{figure*}[!htp]
\centering
\begin{subfigure}[t]{0.48\linewidth}
\centering
\includegraphics[width=\linewidth]{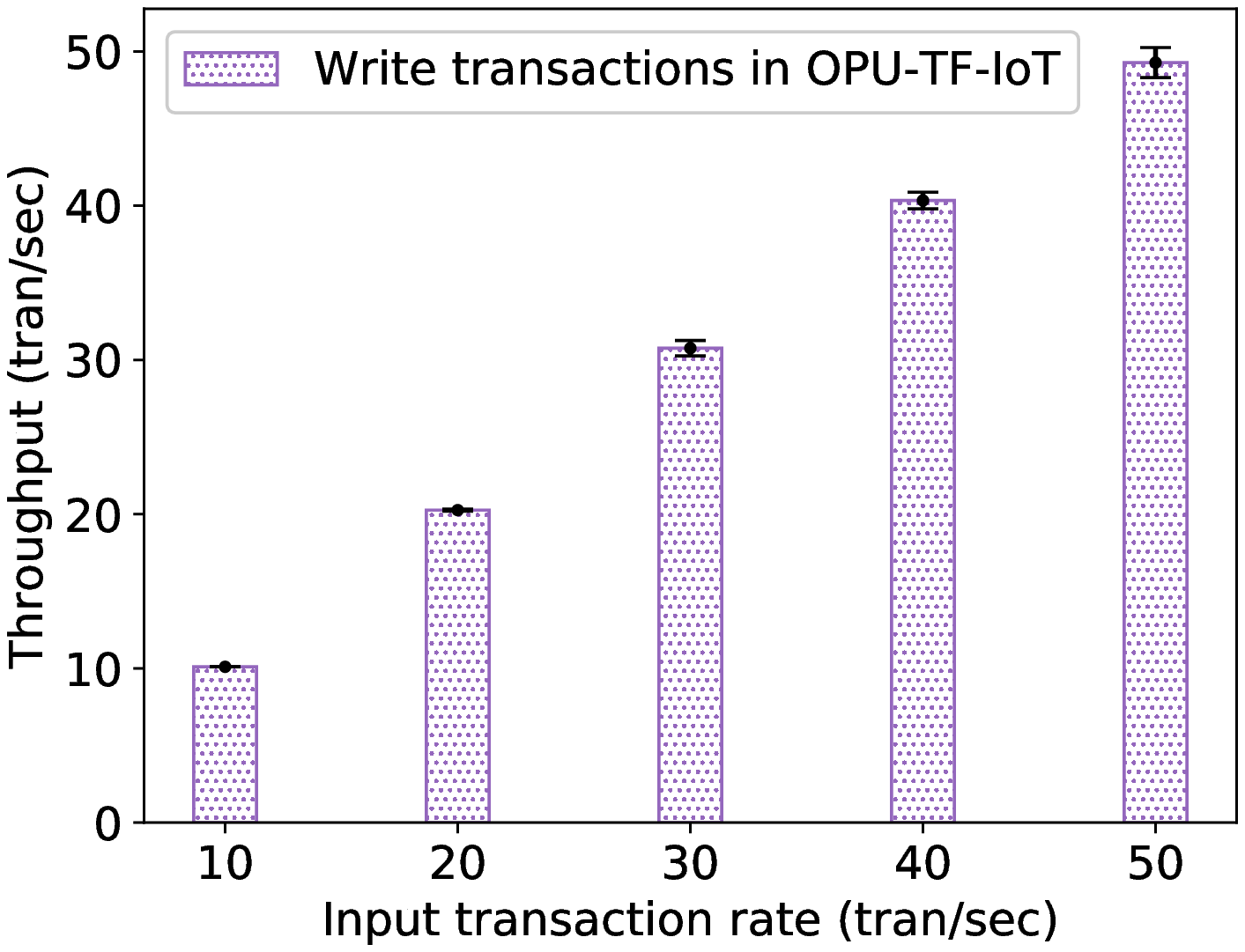}
\caption{Throughput}
\label{th_for_write}
\end{subfigure}
\begin{subfigure}[t]{0.48\linewidth}
\centering
\includegraphics[width=\linewidth]{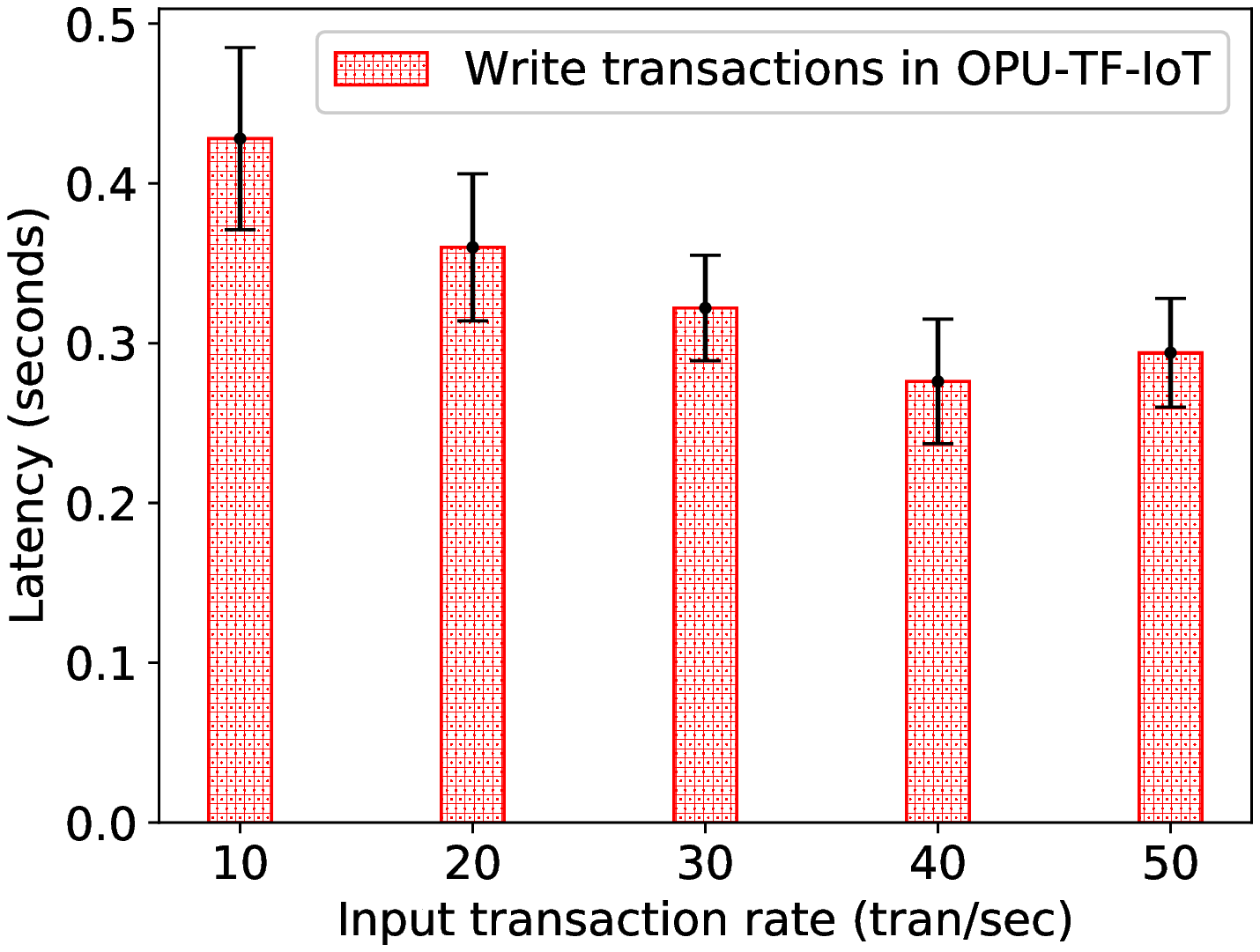}
\caption{Latency}
\label{lat_for_write}
\end{subfigure}
\begin{subfigure}[t]{0.48\linewidth}
\centering
\includegraphics[width=\linewidth]{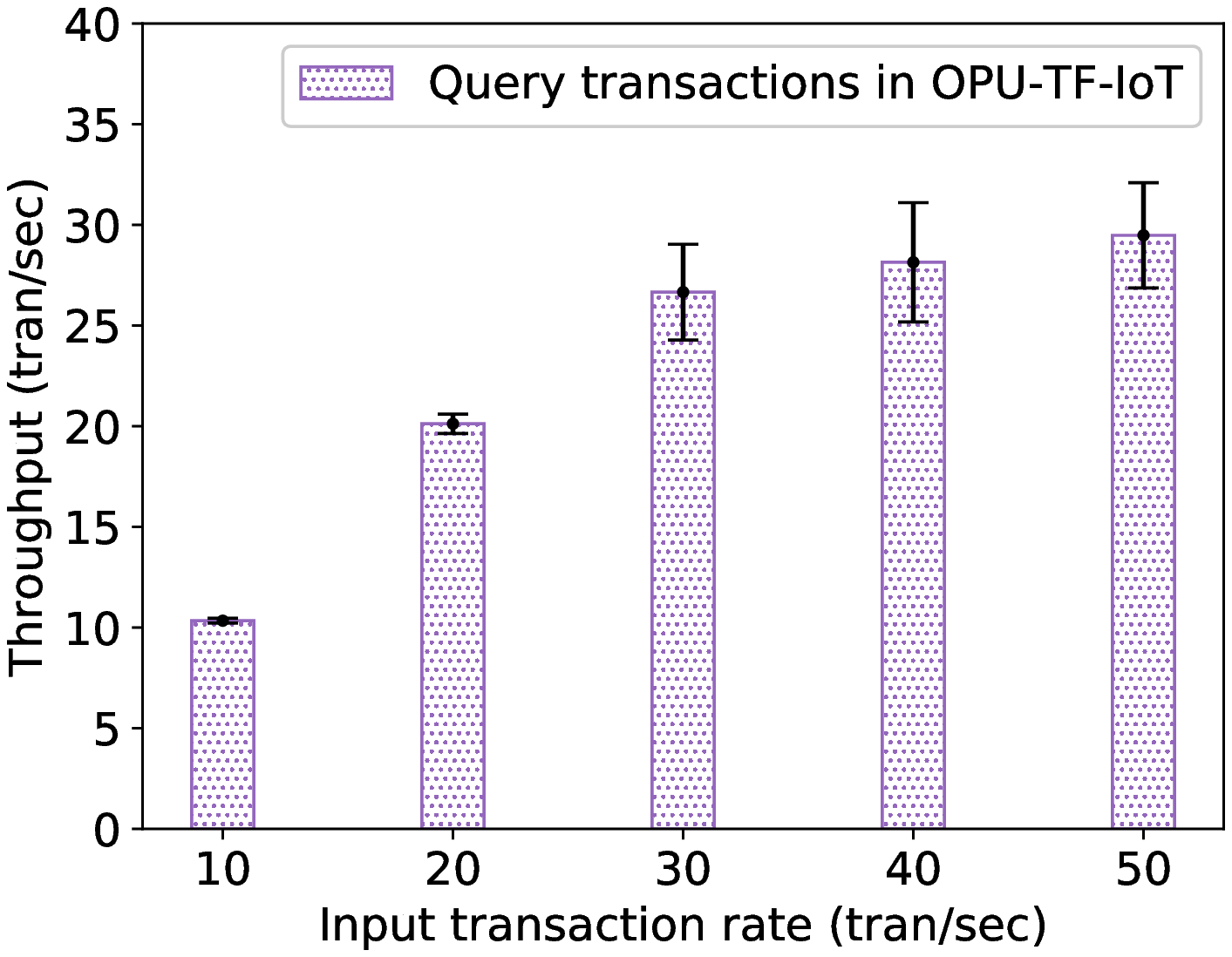}
\caption{Throughput}
\label{th_for_query}
\end{subfigure}
\begin{subfigure}[t]{0.48\linewidth}
\centering
\includegraphics[width=\linewidth]{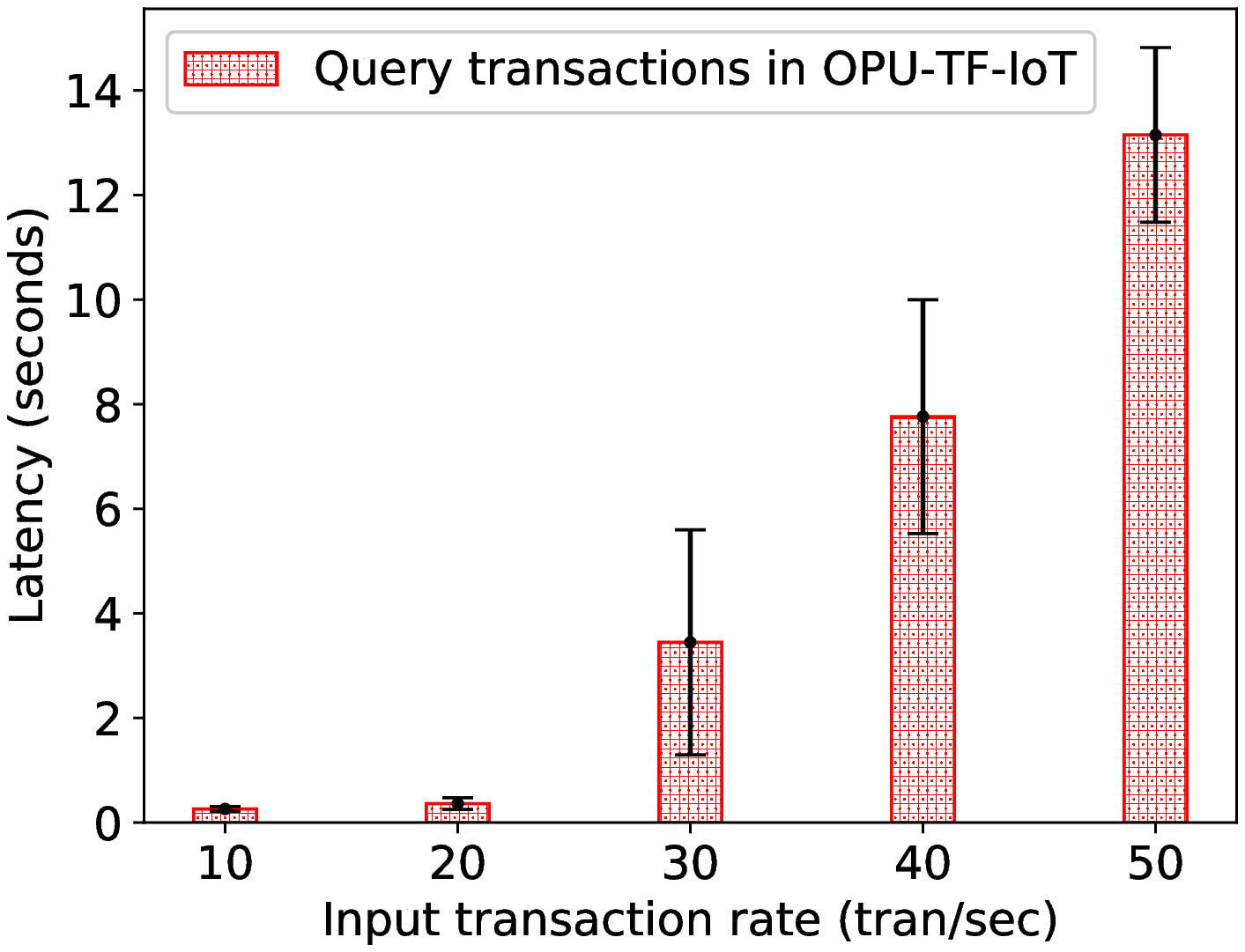}
\caption{Latency}
\label{lat_for_query}
\end{subfigure}
\caption{Evaluation of differential privacy budget verification mechanism in OPU-TF-IoT. The results are within 95\% of confidence interval. Here, we did not compare \cite{dpint2006} and \cite{trackbudget} because the blockchain implementation of both these works is missing. However, the analysis of differential privacy for the mentioned references and proposed work is given in Fig. \ref{fig:eval_and_com_of_priv_util_trad_off} and Table \ref{tab:com_of_tot_budg}.}
\label{fig:thlat}
\end{figure*}

\subsection{Experimental Setup} 
\label{subsec:exs_setup}
\paragraph{Software and Hardware configuration}
To simulate the environment for evaluation, we consider a general network in an IoT scenario which consists of a single curator \textbf{C}, a set of  data owners $\textbf{O} = \{O_{1}, O_{2}, O_{3}...O_{n} \}$, and a set of 10 data users $\textbf{U} = \{U_{1}, U_{2}, U_{3}...U_{k}\}$ where $k = 10$. The curator collects data from the set of data owners \textbf{O} which are IoT devices such as cellular phone, or home appliances. Similarly, we use Hyperledger fabric to establish a blockchain network which consists of two organizations namely the curator and one of the data users. One data user is considered for simplicity which can be easily extended to multiple data users. Furthermore, each organization has one peer and a Couch database connected through a single channel called \textit{mychannel} \cite{hyper}. Moreover, a single smart contract is installed on each of the peer. For data table $T_{n, m}$, we use the free available adult dataset from \cite{adult_data} which consists of 32K records ($n = 32K$) with 16 attributes ($m = 16$). As a result, it is assumed that the data is associated with the set of data owners $\textbf{O} = \{O_{1}, O_{2}, O_{3}...O_{n} \}$ where $n = 32K$. Random queries $\{q_{1}, q_{2}, q_{3}… q_{k}\}$ with $k = 10$ are simulated whereas each query $q_{i}$ is randomly generated which asks a numeric value according to $F \in \{Count, Average, Maximum, Minimum\}$. \\
The required accuracies $A^{i}_{req}$ of the queries are simulated according to $\{0.99, 0.98, 0.96, 0.96, 0.95, 0.93, 0.99, 0.98, 0.95, 0.97\}$. For query type $N$, we consider that the first five queries have type $N = 1$ whereas the last five queries have type $N = 0$. To differentiate the query types $N$, queries with type $N = 0$ are configured with smaller number of requested attributes (a portion of the dataset) in the predicate than the queries with type $N = 1$. Similarly, we take the total privacy budget $\epsilon_{t} = 8$ whereas $\epsilon_{def}$ is varied from 0.1 to 1 with the increment of 0.1. Furthermore, the decrement factor $\eta$ is varied according to $\{0.0005, 0.005, 0.05\}$ and the tolerance factor is taken as $\tau = 0.02$. \\ 
The proposed heuristic search algorithm is implemented in Python to perform the selection of suitable privacy budget whereas the proposed privacy budget verification mechanism is implemented through Hyperledger fabric. Moreover, Hyperledger fabric is used as the target SUT (software under test) with the SDK version 1.4.11. We use Caliper version 0.4.0 for evaluation of the target SUT \cite{caliper}. Similarly, we use Ubuntu-18 64-bit operating system which is installed along with Windows 10 using Oracle VM VirtualBox. The hardware configuration of the system includes Intel(R)Core(TM) i5-8250U CPU @ 1.6 GHz processor with 8 GB of installed physical memory.

\begin{table*}[!htp]
\caption{Evaluation of the impact of $\eta$ on the performance of OPU-TF-IoT with $\tau = 0.02$ and $\eta \in \{0.005, 0.005, 0.05\}$ where Count, Avg, Max, and Min represent Count, Average, Maximum, and Minimum queries, respectively.}

	    \begin{subtable}[h]{0.30\linewidth}
        \centering
        \begin{tabular}{|c|c|c|c|c|}
\hline 
\multirow{2}{*}{$\epsilon_{def}$} & \multicolumn{4}{c|}{\textbf{No of satisfied data users}} \\
\cline{2-5}       
 & Count & Avg & Max & Min   \\ 
\hline
 0.1 & 6 & 10 & 10 & 10  \\
\hline
 0.2 & 8 & 10 & 10 & 10   \\
\hline
 0.3 & 10 & 10 & 10 & 10   \\
\hline
 0.4 & 10 & 10 & 10 & 10  \\
\hline
 0.5 & 10 & 10 & 10 & 10   \\
\hline
 0.6 & 10 & 10 & 10 & 10   \\
\hline
 0.7 & 10 & 10 & 10 & 10   \\
\hline
 0.8 & 10 & 10 & 10 & 10   \\
\hline
 0.9 & 10 & 10 & 10 & 10   \\
\hline
 1 & 10 & 10 & 10 & 10   \\
\hline
        \end{tabular}
\vspace{0.2cm}        
        \caption{$\eta = 0.0005$}
        \label{tab:tab_0005}
     \end{subtable}    
    \hfill
    \begin{subtable}[h]{0.30\linewidth}
        \centering
        \begin{tabular}{|c|c|c|c|c|}
\hline        
\multirow{2}{*}{$\epsilon_{def}$} & \multicolumn{4}{c|}{\textbf{No of satisfied data users}} \\
\cline{2-5}
 & Count & Avg & Max & Min  \\ 
\hline
 0.1 & 5 & 5 & 5 & 8  \\
\hline
 0.2 & 9 & 5 & 5 & 8   \\
\hline
 0.3 & 10 & 5 & 5 & 8  \\
\hline
 0.4 & 10 & 5 & 5 & 8  \\
\hline
 0.5 & 10 & 5 & 5 & 7   \\
\hline
 0.6 & 10 & 5 & 5 & 8   \\
\hline
 0.7 & 10 & 5 & 5 & 7   \\
\hline
 0.8 & 10 & 5 & 5 & 8   \\
\hline
 0.9 & 10 & 5 & 5 & 8   \\
\hline
 1 & 10 & 5 & 5 & 8   \\
\hline
        \end{tabular}
\vspace{0.2cm}        
        \caption{$\eta = 0.005$}
        \label{tab:tab_005}
     \end{subtable}
    \hfill
    \begin{subtable}[h]{0.30\linewidth}
        \centering
        \begin{tabular}{|c|c|c|c|c|}
\hline
\multirow{2}{*}{$\epsilon_{def}$} & \multicolumn{4}{c|}{\textbf{No of satisfied data users}} \\
\cline{2-5}
 & Count & Avg & Max & Min  \\ 
\hline
 0.1 & 4 & 4 & 5 & 5  \\
\hline
 0.2 & 7 & 4 & 5 & 5   \\
\hline
 0.3 & 8 & 4 & 5 & 5   \\
\hline
 0.4 & 10 & 4 & 5 & 5  \\
\hline
 0.5 & 9 & 4 & 5 & 5   \\
\hline
 0.6 & 9 & 4 & 5 & 5   \\
\hline
 0.7 & 9 & 4 & 5 & 5   \\
\hline
 0.8 & 10 & 4 & 5 & 5   \\
\hline
 0.9 & 10 & 4 & 5 & 5   \\
\hline
 1 & 10 & 4 & 5 & 5   \\
\hline
       \end{tabular}
       \vspace{0.2cm}
       \caption{$\eta = 0.05$}
       \label{tab:tab_05}
    \end{subtable}     
     
     \label{tab:impact_of_eta}
\end{table*}

\paragraph{Benchmark configuration}
\label{bench}
The benchmark configuration of the Caliper tool consists of two rounds which are initialization of the ledger and querying the ledger. In the first round, a test with five workers is simulated which sends write transactions with a varying transaction rate from 10 tran/sec to 50 tran/sec to the Hyperledger fabric SUT. The initialization of write transaction with the given parameters is performed through the SUTAdapter as shown in Fig. \ref{fig:write_tran_init}. In the second round, the application sends query transaction (as shown in Fig. \ref{fig:flow_dia+tran_resp}(b)) which is evaluated by peers on the ledger to generate query responses. The simulation results obtained from the experimental setup are presented in the next section.
%\vspace{-5mm}
\subsection{Results and Discussion}
\label{subsec:res_and_dis}
\subsubsection{Optimized privacy-utility trade-off} 
\label{subsubsec:opt_prive_util_trad}
The privacy-utility trade-off comparison is evaluated and presented in Fig. \ref{fig:eval_and_com_of_priv_util_trad_off}. It can be seen from Fig. \ref{fig:eval_and_com_of_priv_util_trad_off} that the BST-DP of \cite{trackbudget} and standard DP of \cite{dpint2006} use a flat allocation of privacy budget for the incoming queries due to which the accumulation of the privacy budget shows a linear increase. In contrast, the accumulation of privacy budget for the OPU-TF-IoT shows variation as we go from left to right. The reason is that OPU-TF-IoT adjusts the privacy budget according to the accuracy requirements of the data users. As a result, it can be seen from Fig. \ref{fig:eval_and_com_of_priv_util_trad_off} that the accumulated privacy budget in OPU-TF-IoT is less than that for the other two approaches for all four query types, i.e., it is (a) 3.12 vs 4.5 and 5 for count, (b) 2.01 vs 4.5 and 5 for average, (c) 2 vs 4.5 and 5 for maximum, and (d) 2.03 vs 4.5 and 5 for minimum queries for OPU-TF-IoT, BST-DP, and standard DP, respectively. Consequently, the OPU-TF-IoT saves the privacy budget by avoiding its waste due to flat allocation of BST-DP by 30.6\%, 55.3\%, 55.5\%, and 54.8\% for count, average, maximum, and minimum queries, respectively as shown in Fig. \ref{fig:eval_and_com_of_priv_util_trad_off}. Similarly, according to Fig. \ref{fig:eval_and_com_of_priv_util_trad_off}, OPU-TF-IoT saves the privacy budget against the standard DP by 37.6\%, 59.8\%, 60\%, and 59.4\% for count, average, maximum, and minimum queries, respectively. \\
The proposed OPU-TF-IoT and state-of-the-art BST-DP reuse the privacy budget for repeated queries which saves the privacy budget as shown in Fig. \ref{fig:eval_and_com_of_priv_util_trad_off}. However, it is evident from Fig. \ref{fig:eval_and_com_of_priv_util_trad_off} that the OPU-TF-IoT outperforms BST-DP. The reason is that BST-DP uses flat allocation scheme for non-repeated queries whereas OPU-TF-IoT uses adjustment of privacy budget according to the accuracy requirements of the data users. Consequently, the utilization of the privacy budget is further improved. Table \ref{tab:com_of_tot_budg} presents the comprehensive comparison of the total privacy budget utilization for different query types. It is evident from the Table \ref{tab:com_of_tot_budg} that for all values of $\epsilon_{def}$, the total privacy budget utilization in OPU-TF-IoT is less than the BST-DP and standard DP for all types of queries which witnesses the improvement of the proposed approach in the utilization and saving of privacy budget. \\
Consequently, it can be deduced from the analysis of the results in Fig. \ref{fig:eval_and_com_of_priv_util_trad_off} and Table \ref{tab:com_of_tot_budg} that the OPU-TF-IoT achieves optimized privacy-utility trade-off by adjusting the privacy budget according to the accuracy requirements of the data users. Furthermore, the data users are satisfied whereas the waste of privacy budget due to flat allocation is avoided which is then utilized for other queries. In this way, the OPU-TF-IoT enables the data curator to answer more queries than the BST-DP and standard DP. 
\subsubsection{Verification of utilization of the privacy budget}
\label{subsubsec:ver_of_priv_bgt}
In this part, we evaluate the privacy budget verification mechanism of OPU-TF-IoT. For this reason, the output parameters of the algorithm \ref{alg:heu_search}, i.e., F, N, $\epsilon$, and A are passed to the submitTransaction function of the client application of Hyperledger fabric as shown in Fig. \ref{fig:write_tran_init}. The submitTransaction function initializes the write transaction which is then used to write the contractArguments to the blockchain ledger. Furthermore, the blockchain network is then evaluated for processing of Init (write) and query transactions. In the proposed experimental setting, throughput and latency of transactions are evaluated to study the maximum processing capacity and latency of transactions of SUT. The results are presented in Fig. \ref{fig:thlat}. It is evident from the results in Fig. \ref{fig:thlat}, that the throughput increases for both write and query transactions, respectively. The reason is that the range of input transaction rate is within the processing capacity of the SUT. Therefore, according to Fig. \ref{fig:thlat}(a) and \ref{fig:thlat}(c), more input transactions in the unit time results in higher throughput. The maximum throughput of 50 and 30 tran/sec are obtained for write and query transactions, respectively.  \\
Similarly, the evaluation of latency is shown in Fig. \ref{fig:thlat}(b) and \ref{fig:thlat}(d). According to Fig. \ref{fig:thlat}(b), for write transactions, the maximum processing capacity reaches for input transactions rate of 40 tran/sec. Therefore, increasing the input transaction rate beyond this point shows increase in the latency of write transactions. In contrast, according to Fig. \ref{fig:thlat}(d), for query transactions, a steep increase beyond 20 tran/sec is detected which shows that the SUT reaches its maximum capacity of transactions processing. As a result, increasing the input transaction rate beyond this point results in abrupt increase in the latency. \\
The results in Fig. \ref{fig:thlat} suggest that the privacy budget verification mechanism of OPU-TF-IoT is suitable for practical scenarios in IoT. The reason is that in the current setting, it achieves a maximum throughput of 50 tran/sec. Similarly, the maximum latency in the current setting is around 12 sec for 50 tran/sec of input transaction rate which is again feasible in practical scenarios. As a result, the privacy budget verification mechanism of OPU-TF-IoT enables the data owners to verify the data sharing activities which increases the transparency of the system. 

\subsubsection{Impact of $\epsilon_{def}$ and $\eta$ on the performance of OPU-TF-IoT}
\label{subsubsec:imp_of_tel_and_decfact}

In this section, we evaluate the impact of $\epsilon_{def}$ and $\eta$ on the number of satisfied data users in OPU-TF-IoT. Table \ref{tab:impact_of_eta} presents the number of satisfied data users as a function of $\epsilon_{def}$ and $\eta$. It is evident from the results that a smaller value of $\eta$ increases the number of satisfied data users. For instance, for $\eta = 0.0005$, the number of satisfied data users is 100\% for all query types except the two cells in the count column as shown in Table \ref{tab:tab_0005}. The reason is that a smaller $\eta$ increments the $\epsilon_{def}$ by a small fraction which enables the curator to find a suitable privacy budget $\epsilon_{sut}$. In contrast, both $\eta = 0.005$ and $\eta = 0.05$ result in lower number of satisfied data users as shown in Tables \ref{tab:tab_005} and \ref{tab:tab_05}, respectively. The reason is that OPU-IT-IoT cannot find a suitable adjusted value of $\epsilon_{sut}$ through the gradual decrement of $\epsilon_{def}$ which is not desired. \\
Similarly, the number of satisfied data users vary with the selection of $\epsilon_{def}$. For example, the results in Table \ref{tab:tab_0005} indicate that the data curator should select $\epsilon_{def} = 0.3$ (row 3 of Table \ref{tab:tab_0005}) instead of 0.1 and 0.2 (rows 1 and 2 of Table \ref{tab:tab_0005}, respectively) to avoid the decrease in the number of satisfied data users. The reason is that if the curator selects a smaller $\epsilon_{def}$ then the data users with high accuracy requirements will not be satisfied. Therefore, the curator in the proposed work uses algorithm \ref{alg:sel_of_ep_and_eta} to keep track of the number of satisfied data users and change the $\epsilon_{def}$ and $\eta$ accordingly. In this way, OPU-TF-IoT increases the number of satisfied data user and avoid the waste of privacy budget at the same time. \\
Finally, from the evaluation results, it is evident that the proposed OUP-TF-IoT outperforms the state-of-the-art BST-DP of \cite{trackbudget} and standard DP of \cite{dpint2006} in terms of optimized privacy-utility trade-off. More specifically, OPU-TF-IoT avoids the waste of privacy budget, increases the number of satisfied data users, and enable the data owners to verify their privacy preservation level by making the data sharing activities transparent and accessible. 
%\vspace{-5mm}
\section{Conclusion} 
\label{sec:con}
In this work, we proposed an optimized privacy-utility trade-off framework (OPU-TF-IIoT) for IoT-based applications. Differential privacy has been adopted to share the data in a privacy preserving manner. Similarly, to optimize the privacy-utility trade-off, we considered the population or dataset size along the query. Furthermore, an algorithm called heuristic search is proposed to adjust the privacy budget according to the accuracy requirements of the data users. Moreover, to avoid the risk of privacy leakage due to central processing of the data, a verification mechanism is also designed through Hyperledger fabric. It was found that the proposed OPU-TF-IoT outperforms the state-of-the-art standard differential privacy of \cite{dpint2006}, and BST-DP of \cite{trackbudget} in terms of optimal privacy-utility trade-off. Finally, it was also validated through the results that the proposed work can be implemented using a cloud server and the transaction processing rate of the Hyperledger fabric is also feasible. Consequently, it enables to share the data in more efficient manner by avoiding the waste of privacy budget, increase the number of satisfied data users, and making the data sharing events transparent to the data owners.  	
\ifCLASSOPTIONcaptionsoff
  \newpage
\fi
%\vspace{-5mm}
\bibliographystyle{IEEEtran}
%\bibliography{IEEEabrv,problem2changed}
\bibliography{main.bbl}

\begin{thebibliography}{10}
\providecommand{\url}[1]{#1}
\csname url@samestyle\endcsname
\providecommand{\newblock}{\relax}
\providecommand{\bibinfo}[2]{#2}
\providecommand{\BIBentrySTDinterwordspacing}{\spaceskip=0pt\relax}
\providecommand{\BIBentryALTinterwordstretchfactor}{4}
\providecommand{\BIBentryALTinterwordspacing}{\spaceskip=\fontdimen2\font plus
\BIBentryALTinterwordstretchfactor\fontdimen3\font minus
  \fontdimen4\font\relax}
\providecommand{\BIBforeignlanguage}[2]{{%
\expandafter\ifx\csname l@#1\endcsname\relax
\typeout{** WARNING: IEEEtran.bst: No hyphenation pattern has been}%
\typeout{** loaded for the language `#1'. Using the pattern for}%
\typeout{** the default language instead.}%
\else
\language=\csname l@#1\endcsname
\fi
#2}}
\providecommand{\BIBdecl}{\relax}
\BIBdecl

\bibitem{IoTsurvey}
M.~Stoyanova, Y.~Nikoloudakis, S.~Panagiotakis, E.~Pallis, and E.~K. Markakis,
  ``A survey on the internet of things {IoT} forensics: Challenges, approaches,
  and open issues,'' \emph{IEEE Communications Surveys Tutorials}, vol.~22,
  no.~2, pp. 1191--1221, 2020.

\bibitem{IoTforsmarthealth}
S.~Raj, ``An efficient {IoT}-based platform for remote real-time cardiac
  activity monitoring,'' \emph{IEEE Transactions on Consumer Electronics},
  vol.~66, no.~2, pp. 106--114, 2020.

\bibitem{IoTforsmartfact}
D.~A. Chekired, L.~Khoukhi, and H.~T. Mouftah, ``Industrial {IoT} data
  scheduling based on hierarchical fog computing: A key for enabling smart
  factory,'' \emph{IEEE Transactions on Industrial Informatics}, vol.~14,
  no.~10, pp. 4590--4602, 2018.

\bibitem{IoTforsmarttransport}
F.~Zhu, Y.~Lv, Y.~Chen, X.~Wang, G.~Xiong, and F.-Y. Wang, ``Parallel
  transportation systems: Toward {IoT}-enabled smart urban traffic control and
  management,'' \emph{IEEE Transactions on Intelligent Transportation Systems},
  vol.~21, no.~10, pp. 4063--4071, 2020.

\bibitem{IoTinsmartcity}
F.~Cirillo, D.~Gómez, L.~Diez, I.~Elicegui~Maestro, T.~B.~J. Gilbert, and
  R.~Akhavan, ``Smart city {IoT} services creation through large-scale
  collaboration,'' \emph{IEEE Internet of Things Journal}, vol.~7, no.~6, pp.
  5267--5275, 2020.

\bibitem{IoTuserpriv1}
S.-C. Cha, T.-Y. Hsu, Y.~Xiang, and K.-H. Yeh, ``Privacy enhancing technologies
  in the internet of things: Perspectives and challenges,'' \emph{IEEE Internet
  of Things Journal}, vol.~6, no.~2, pp. 2159--2187, 2019.

\bibitem{IoTuserpriv2}
M.~A. Lisovich, D.~K. Mulligan, and S.~B. Wicker, ``Inferring personal
  information from demand-response systems,'' \emph{IEEE Security Privacy},
  vol.~8, no.~1, pp. 11--20, 2010.

\bibitem{IoTuserpriv3}
W.~Lin, X.~Zhang, L.~Qi, W.~Li, S.~Li, V.~S. Sheng, and S.~Nepal,
  ``Location-aware service recommendations with privacy-preservation in the
  internet of things,'' \emph{IEEE Transactions on Computational Social
  Systems}, vol.~8, no.~1, pp. 227--235, 2021.

\bibitem{differentialpubsurvey}
T.~Zhu, G.~Li, W.~Zhou, and S.~Y. Philip, ``Differentially private data
  publishing and analysis: A survey,'' \emph{IEEE Transactions on Knowledge and
  Data Engineering}, vol.~29, no.~8, pp. 1619--1638, 2017.

\bibitem{Dwork2008}
C.~Dwork, ``Differential privacy: A survey of results,'' in \emph{Theory and
  Applications of Models of Computation}, M.~Agrawal, D.~Du, Z.~Duan, and
  A.~Li, Eds.\hskip 1em plus 0.5em minus 0.4em\relax Berlin, Heidelberg:
  Springer Berlin Heidelberg, 2008, pp. 1--19.

\bibitem{RLbasqueropt}
Y.~Jiang, K.~Zhang, Y.~Qian, and L.~Zhou, ``Reinforcement-learning-based query
  optimization in differentially private {IoT} data publishing,'' \emph{IEEE
  Internet of Things Journal}, vol.~8, no.~14, pp. 11\,163--11\,176, 2021.

\bibitem{gambasedpriv_utility}
X.~Wu, T.~Wu, M.~Khan, Q.~Ni, and W.~Dou, ``Game theory based correlated
  privacy preserving analysis in big data,'' \emph{IEEE Transactions on Big
  Data}, vol.~7, no.~4, pp. 643--656, 2021.

\bibitem{islamtransparency}
M.~Islam, M.~H. Rehmani, and J.~Chen, ``Transparency-privacy trade-off in
  blockchain-based supply chain in industrial internet of things,'' in
  \emph{2021 IEEE 23rd Int Conf on High Performance Computing and
  Communications; 7th Int Conf on Data Science and Systems; 19th Int Conf on
  Smart City; 7th Int Conf on Dependability in Sensor, Cloud and Big Data
  Systems and Application (HPCC/DSS/SmartCity/DependSys)}, 2021, pp.
  1123--1130.

\bibitem{dist-class-priv-acc}
L.~{Xu}, C.~{Jiang}, Y.~{Qian}, J.~{Li}, Y.~{Zhao}, and Y.~{Ren},
  ``Privacy-accuracy trade-off in differentially-private distributed
  classification: A game theoretical approach,'' \emph{IEEE Transactions on Big
  Data}, pp. 1--1, 2017.

\bibitem{privacy-dist}
B.~{Rassouli} and D.~{Gündüz}, ``Optimal utility-privacy trade-off with total
  variation distance as a privacy measure,'' \emph{IEEE Transactions on
  Information Forensics and Security}, vol.~15, pp. 594--603, 2020.

\bibitem{adoptive_person_DP}
B.~Niu, Y.~Chen, B.~Wang, Z.~Wang, F.~Li, and J.~Cao, ``Adapdp: Adaptive
  personalized differential privacy,'' in \emph{IEEE INFOCOM - IEEE Conference
  on Computer Communications}, 2021, pp. 1--10.

\bibitem{util_awar_gen_framework}
H.~Jiang, M.~Wang, P.~Zhao, Z.~Xiao, and S.~Dustdar, ``A utility-aware general
  framework with quantifiable privacy preservation for destination prediction
  in lbss,'' \emph{IEEE/ACM Transactions on Networking}, vol.~29, no.~5, pp.
  2228--2241, 2021.

\bibitem{trackbudget}
Y.~{Zhao}, J.~{Zhao}, J.~{Kang}, Z.~{Zhang}, D.~{Niyato}, S.~{Shi}, and K.~Y.
  {Lam}, ``A blockchain-based approach for saving and tracking
  differential-privacy cost,'' \emph{IEEE Internet of Things Journal}, pp.
  1--1, 2021.

\bibitem{dp_pop_size}
A.~Friedman and A.~Schuster, ``Data mining with differential privacy,'' in
  \emph{Proceedings of the 16th ACM SIGKDD International Conference on
  Knowledge Discovery and Data Mining}, ser. KDD '10.\hskip 1em plus 0.5em
  minus 0.4em\relax New York, NY, USA: Association for Computing Machinery,
  2010, p. 493–502.

\bibitem{contract-theoratic}
L.~{Xu}, C.~{Jiang}, Y.~{Chen}, Y.~{Ren}, and K.~J.~R. {Liu}, ``Privacy or
  utility in data collection? a contract theoretic approach,'' \emph{IEEE
  Journal of Selected Topics in Signal Processing}, vol.~9, no.~7, pp.
  1256--1269, 2015.

\bibitem{ppass2021}
\BIBentryALTinterwordspacing
M.~Chamikara, P.~Bertok, I.~Khalil, D.~Liu, and S.~Camtepe, ``Ppaas: Privacy
  preservation as a service,'' \emph{Computer Communications}, vol. 173, pp.
  192--205, 2021. [Online]. Available:
  \url{https://www.sciencedirect.com/science/article/pii/S0140366421001420}
\BIBentrySTDinterwordspacing

\bibitem{dwork2014algorithmic}
C.~Dwork, A.~Roth \emph{et~al.}, ``The algorithmic foundations of differential
  privacy.'' \emph{Foundations and Trends in Theoretical Computer Science},
  vol.~9, no. 3-4, pp. 211--407, 2014.

\bibitem{nakamoto2008}
S.~Nakamoto and A.~Bitcoin, ``A peer-to-peer electronic cash system,''
  \emph{Bitcoin.--URL: https://bitcoin. org/bitcoin. pdf}, 2008.

\bibitem{anomalyinBC}
M.~U. Hassan, M.~H. Rehmani, and J.~Chen, ``Anomaly detection in blockchain
  networks: A comprehensive survey,'' \emph{IEEE Communications Surveys and
  Tutorials}, pp. 1--1, 2022.

\bibitem{compriot2020}
J.~Sengupta, S.~Ruj, and S.~D. Bit, ``A comprehensive survey on attacks,
  security issues and blockchain solutions for {IoT} and {IIoT},''
  \emph{Journal of Network and Computer Applications}, vol. 149, p. 102481,
  2020.

\bibitem{hyper}
``Hyperledger-fabricdocs documentation,''
  \url{https://hyperledger-fabric.readthedocs.io/en/release-2.2/}, accessed:
  2021-02-20.

\bibitem{relterror}
X.~Xiao, G.~Bender, M.~Hay, and J.~Gehrke, ``Ireduct: Differential privacy with
  reduced relative errors,'' in \emph{Proceedings of the ACM SIGMOD
  International Conference on Management of Data}, ser. SIGMOD '11.\hskip 1em
  plus 0.5em minus 0.4em\relax New York, NY, USA: Association for Computing
  Machinery, 2011, p. 229–240.

\bibitem{dpint2006}
C.~Dwork, ``Differential privacy,'' in \emph{Automata, Languages and
  Programming}, M.~Bugliesi, B.~Preneel, V.~Sassone, and I.~Wegener, Eds.\hskip
  1em plus 0.5em minus 0.4em\relax Berlin, Heidelberg: Springer Berlin
  Heidelberg, 2006, pp. 1--12.

\bibitem{adult_data}
``Adult data,'' \url{https://archive.ics.uci.edu/ml/datasets/Adult}, accessed:
  2022-02-10.

\bibitem{caliper}
``Hyperledger caliper,'' \url{https://www.hyperledger.org/use/caliper},
  accessed: 2021-02-20.

\end{thebibliography}
\end{document}